\newcommand{\figref}[1]{Fig.~\ref{#1}}

\newcommand{\equref}[1]{Eq.~(\ref{#1})}

\documentclass[lettersize,journal]{IEEEtran}
\usepackage{amssymb}
\usepackage{amsmath}
\usepackage{graphicx}
\usepackage{multirow}
\usepackage{algorithm}
\usepackage{algpseudocode}
\usepackage{epstopdf}
\usepackage{amsmath}
\usepackage{algorithmicx}
\usepackage{algpseudocode}
\usepackage{array}
\usepackage{color}
\usepackage{xcolor}
\usepackage{graphics}
\usepackage{multirow}
\usepackage{setspace}   
\usepackage{enumitem}
\usepackage{epstopdf}
\usepackage{pdfpages}
\usepackage{amssymb}
\usepackage{graphicx}
\usepackage{hyperref}
\usepackage{booktabs}
\usepackage{makecell}
\usepackage{float}
\hypersetup{
	colorlinks=true,   
	linkcolor=teal,     
	citecolor=magenta,   
	filecolor=mycustompurple,   
	urlcolor=red    
}
\makeatletter

\newcommand{\Rmnum}[1]{\expandafter\@slowromancap\romannumeral #1@}
\newtheorem{remark}{Remark}
\makeatother
\usepackage{balance}

\begin{document}
	\title{RIS-Empowered OTFS Modulation With Faster-than-Nyquist Signaling in High-Mobility Wireless Communications}
	
	\author{Chaorong Zhang,\
	Benjamin K. Ng, \
	Hui Xu,
	Chan-Tong Lam, and\
	Halim Yanikomeroglu
	\thanks{\textsl{(Corresponding author: Benjamin K. Ng.)}
		}}

	\maketitle

\begin{abstract}
Faster-than-Nyquist (FTN) signaling offers a promising path to enhance the spectral efficiency (SE) of orthogonal time frequency space (OTFS) modulation in high-mobility scenarios. 
However, the inherent non-orthogonal pulse shaping induces severe structured interference and colored noise, leading to reliability degradation and error floors. 
This paper proposes a reconfigurable intelligent surface (RIS)-empowered OTFS-FTN framework, demonstrating that the RIS acts as a critical enabler rather than a generic SNR booster. 
By strategically configuring RIS phases, we reshape the cascaded delay-Doppler (DD) channel to mitigate FTN-induced impairments, thereby enhancing reliability and relaxing power amplifier (PA) back-off constraints. 
We establish a unified DD-domain model explicitly accounting for FTN-induced colored noise and derive analytical expressions for the average frame error rate (AFER) upper bound, SE, peak-to-average power ratio (PAPR), and input back-off (IBO). 
Moreover, a low-complexity RIS phase design is proposed, with quantified computational and control overheads. 
Extensive simulations against comprehensive baselines, including Nyquist OTFS and power-compensated benchmarks, validate that the proposed framework achieves a superior SE-reliability-IBO trade-off. 
Results show up to around 10 dB reduction in required SNR for equivalent SE and AFER, confirming gains that cannot be replicated by simply optimizing simpler systems.
\end{abstract}


\section{Introduction}

\subsection{Orthogonal Time Frequency Space}
As emerging technologies like the Internet of Things (IoT), Vehicle-to-Everything (V2X), and low-Earth-orbit (LEO) satellite communications continue to advance, there is growing focus on improving the quality of mobile wireless communications \cite{ref1}-\cite{ref3}. 
However, traditional wireless schemes such as the Orthogonal Frequency Division Multiplexing (OFDM) perform poorly in high-mobility scenarios, particularly in terms of error performance, mainly because they lack strong robustness against delay and Doppler effects \cite{ref4}.
In contrast, the Orthogonal Time Frequency Space (OTFS) modulation offers a better solution to this challenge. 
OTFS is an emerging two-dimensional modulation scheme that represents information symbols in the delay–Doppler (DD) domain, rather than in the conventional time–frequency (TF) domain \cite{ref5}-\cite{ref6}. 
By exploiting the quasi-static nature of wireless channels in the DD domain, OTFS achieves a more compact and stable channel representation, enabling all information symbols to experience the full TF diversity of the channel. 
This property makes OTFS particularly robust against severe Doppler spread and delay dispersion, which are common in high-mobility and frequency-selective environments such as high-speed trains, vehicular communications, and LEO satellite links \cite{ref7}-\cite{ref8}. 
Despite this, there remains significant potential to further enhance OTFS's error performance and spectral efficiency under time-varying channel conditions.
Motivated by this, enhancing OTFS in high-mobility regimes calls for jointly improving spectral efficiency and reliability, which naturally leads to combining FTN signaling with controllable propagation technologies under a unified design and analysis framework. 

\subsection{Faster-than-Nyquist}
As one of the famous high-efficiency techniques in wireless communications, the faster-than-Nyquist (FTN) signaling is a promising transmission technique that intentionally compresses the symbol spacing below the limited Nyquist rate in the time domain, thereby increasing the data rate within a given bandwidth \cite{ref9}-\cite{ref10}. 
By introducing a controllable and acceptable level of intentional inter-symbol interference (ISI), FTN enables more symbols to be transmitted over the same spectral resources, effectively enhancing spectral efficiency without expanding the occupied bandwidth. 
Several state-of-the-art works show that a carefully designed FTN system can achieve higher data rates without a proportional degradation in error performance, provided that advanced detection or equalization algorithms are employed \cite{ref11}-\cite{ref15}. 
This property makes FTN particularly attractive for modern and future wireless communication scenarios where spectral resources are scarce, such as high-capacity terrestrial wireless networks, IoT applications, and next-generation satellite systems.
Moreover, FTN can be flexibly combined with emerging modulation techniques like OTFS, leveraging the DD domain processing to mitigate ISI and improve robustness in high-mobility environments.
FTN signaling improves the SE of OTFS by intentionally accelerating the symbol rate. Yet, the resulting non-orthogonal pulse shaping produces structured interference and colored noise after matched filtering and sampling, which fundamentally differs from the white-noise assumption in conventional OTFS receivers. 
However, the effective DD-domain channel becomes less sparse and the receiver must cope with both intensified inter-symbol coupling and noise correlation with FTN application, potentially leading to error floors and sharply increased detection complexity.

\subsection{Reconfigurable Intelligent Surface}
Thus, a novel and popular wireless technique called the Reconfigurable intelligent surfaces (RIS) is introduced in our work.
RIS has recently emerged as a promising enabling technology for beyond fifth-generation (B5G) and sixth-generation (6G) wireless communication systems. 
An RIS typically consists of a two-dimensional (2D) planar array of reconfigurable reflection elements each fabricated from engineered materials capable of manipulating incident electromagnetic waves in a programmable manner \cite{ref16}.
Each reflection element can be individually controlled by an intelligent controller or chip, which receives configuration instructions from the transmitter and adjusts the phase shifts of the reflected signals accordingly \cite{ref17}. 
This unique capability enables RIS to reshape the wireless propagation environment by steering the reflected beams toward desired directions, thereby extending coverage, enhancing received signal power, and improving link reliability.
Unlike conventional active relays such as unmanned aerial vehicles or vehicular relays, RIS operates without radio-frequency (RF) chains, power amplifiers, or active signal regeneration \cite{ref18}. 
As a result, it can achieve high beamforming gains with minimal power consumption, a feature commonly referred to as passive beamforming \cite{ref19}-\cite{ref20}. 
Through intelligent phase shift optimization, RIS can enhance the signal-to-noise ratio (SNR) at the receiver while incurring negligible additional energy cost, making it highly attractive for energy-efficient wireless communications. 
Since FTN causes degradation in bit-error rate (BER) performance and OTFS still fails to achieve sufficiently satisfactory performance in high-mobility scenarios, RIS can be considered to address these issues.
Despite its robustness, RIS-OTFS faces a fundamental SE ceiling imposed by the Nyquist criterion. Although high-order modulations can improve SE, they degrade severely in harsh fading channels. Consequently, existing systems suffer from a rigid robustness-efficiency trade-off, preventing the simultaneous achievement of high data rates and reliable transmission.
In our work, RIS provides benefits beyond simple SNR enhancement. 
When FTN-aided interference dominates, only increasing transmit power is inefficient and limited by power amplifiers (PA) back-off (IBO) and peak-to-average power ratio (PAPR)-induced nonlinearity. 
By configuring phases, RIS can reshape the cascaded channel to improve post-equalization reliability under FTN. 
This reduces the required transmit power and expands the feasible IBO region for a target AFER and SE, demonstrating a non-trivial synergy between RIS and OTFS-FTN beyond a mere combination of two.

\subsection{Related Works}
Recently, several studies investigate the RIS-assisted OTFS (RIS-OTFS) and FTN-assisted OTFS (FTN-OTFS) schemes. 
On the one hand, the pure RIS-OTFS schemes are first proposed and analyzed in \cite{ref21}-\cite{ref25}, where RIS significantly improves the BER performance of OTFS systems. 
In addition, fundamental aspects such as the input–output relationship, RIS phase adjustment design, multiple-input multiple-output (MIMO) system modeling, and channel estimation are thoroughly discussed in these works. 
However, further theoretical investigation of RIS-OTFS is still required, particularly in terms of error performance, spectral efficiency, and the power fluctuation characteristics of modulated signals, in order to provide deeper insights for practical applications. 
Later, \cite{ref26} proposes a novel index modulation-aided RIS-OTFS scheme to enhance spectral efficiency, but it further increases the detection complexity that is already high in OTFS system. 
As another high-efficiency technology, FTN serves as a better alternative to index modulation, offering lower detection complexity while supporting higher data rates.
On the other hand, various FTN-OTFS schemes are proposed and examined in \cite{ref27}-\cite{ref30}. 
While these schemes improve the spectral efficiency of OTFS systems, they often suffer from degraded BER performance. 
In particular, when frequency compression is applied, as in \cite{ref29}-\cite{ref30}, the BER performance further deteriorates due to the increased ISI. 
Plus, \cite{RR_R2} demonstrates a low-complexity transceiver for OTFS-FTN that applies Reverse Cuthill-McKee reordering and Cholesky factorization to the sparse channel matrix. This reduces complexity from cubic to near-linear in frame size, providing a feasible path to deploy the proposed system at scale.
Overall, both of these fields lack sufficient literature to provide a stronger theoretical foundation and more insights, leaving a research gap and potential improvment among the RIS, FTN, and OTFS.

\subsection{Contributions}
	OTFS mitigates Doppler spread in high-mobility scenarios but is limited to Nyquist-rate SE. FTN surpasses this limit via compressed symbol spacing, yet introduces severe ISI that degrades reliability. 
	To address this, we employ RIS. Its passive beamforming gain effectively offsets the SNR penalty incurred by FTN. 
	This synergy supports aggressive FTN compression, unlocking high SE without compromising the reliability of OTFS in challenging channels, thereby overcoming the traditional mobility-SE trade-off.
	In light of the above works and motivated aforementioned challenges in reliability and spectral efficiency, a novel wireless scheme called the RIS-empowered OTFS modulation with FTN signaling (RIS-OTFS-FTN) scheme is proposed and investigated in this paper.
Main contributions of our work are briefly summarized as
\begin{itemize}
	\item \textbf{The RIS-OTFS-FTN scheme is firstly proposed andcomprehensively investigated in this paper.} The input–output relationship in the DD domain is provided, which explicitly characterizes the joint effects of RIS-induced passive beamforming, FTN-induced intentional ISI, and the inherent delay–Doppler channel dynamics. This unified mathematical framework lays the foundation for systematic performance evaluation and facilitates the development of optimized signal processing.
	\item \textbf{We perform a theoretical analysis of the average frame error rate (AFER) for the proposed RIS-OTFS-FTN scheme.} By leveraging the derived DD-domain model, we obtain closed-form FER expressions that capture the influence of key system parameters. These analytical results not only reveal insightful trade-offs between spectral efficiency and reliability but also serve as valuable guidelines for system design in high-mobility scenarios.
	\item \textbf{Since no existing studies on RIS-OTFS systems have provided relevant analytical performance results}, we present theoretical formulations for evaluating spectral efficiency, PAPR, and other aspects of analytical performance. These formulations offer a theoretical foundation and framework with valuable insights, supporting practical applications and further system extensions.
	\item \textbf{A tailored RIS phase adjustment strategy that optimizes the discrete phase shifts of individual reflection elements based on the DD channel state information is designed.} This design aligns reflected signal components constructively at receiver, thus maximizing the effective channel gain. The proposed method accounts for practical hardware constraints by employing quantized phase control, while still achieving near-optimal performance.
	\item \textbf{We present extensive Monte Carlo simulations and theoretical results, along with a series of in-depth performance insights}, including the interplay among frame error rate (FER), spectral efficiency, PAPR, complementary cumulative distribution function (CCDF), IBO, and BER in high-mobility scenarios. Furthermore, to approximate practical high-speed vehicular scenarios in urban environments, the standardized extended vehicular A (EVA) channel model is adopted in the simulations.
\end{itemize}

\subsection{Notation}
\textit{Notation:} 
The superscripts $(\cdot)^{T}$, $(\cdot)^{H}$, and $(\cdot)^{\dagger}$ denote the transpose, Hermitian (conjugate) transpose, and conjugate transpose, respectively, where $(\cdot)^{H} \equiv (\cdot)^{\dagger}$. 
The operator $\mathrm{diag}(\cdot)$ forms a diagonal matrix from its vector argument, and $\mathrm{diag}(\lambda_1,\dots,\lambda_r)$ denotes a diagonal matrix with diagonal entries $\lambda_1,\dots,\lambda_r$. 
The operator $\mathrm{vec}(\cdot)$ stacks the columns of a matrix into a single column vector. 
The notation $\mathbb{I}_n$ denotes the $n \times n$ identity matrix, and $\mathbb{I}_{MN}$ denotes the $(MN) \times (MN)$ identity matrix. 
The set $\mathbb{C}^{m\times n}$ represents the space of $m\times n$ complex-valued matrices, and $\mathbb{C}$ denotes the set of complex numbers. 
The Euclidean norm is denoted by $\|\cdot\|$, and $|\cdot|$ denotes the absolute value of a scalar. 
The operator $\mathrm{rank}(\cdot)$ denotes the rank of a matrix, and $\det(\cdot)$ denotes the determinant. 
The operator $\mathrm{spec}\left( \cdot \right)$ denotes the set of eigenvalues of a matrix, representing its spectrum.
The notation $\lfloor \cdot \rfloor$ is the floor (integer part) operator. 
The circular convolution is denoted by $\circledast$, and $\otimes$ represents the Kronecker product. 
The notation $\delta(\cdot)$ represents the Dirac delta function. 
The real part of a complex number $x$ is denoted by $\Re\{x\}$. 
The expectation operator is denoted by $\mathbb{E}[\cdot]$, and $\Pr(\cdot)$ denotes probability. 
The notation $\mathcal{S}$ represents the codebook set, with $|\mathcal{S}|$ being its cardinality. 
The operator $\log_2(\cdot)$ denotes the base-2 logarithm. 
The notation $z_i \sim \mathcal{CN}(0,\sigma^2)$ indicates that the $i$-th entry follows a zero-mean circularly symmetric complex Gaussian distribution with variance $\sigma^2$.
The exponential function is denoted by both $\exp \!\:(\cdot )$ and $e^{\left( \cdot \right)}$, which, while distinct in notation, convey the identical mathematical concept.
$h\left( \cdot \right)$ is differential entropy.

\begin{figure}
	\centering
	\includegraphics[width=8.5cm,height=4.3cm]{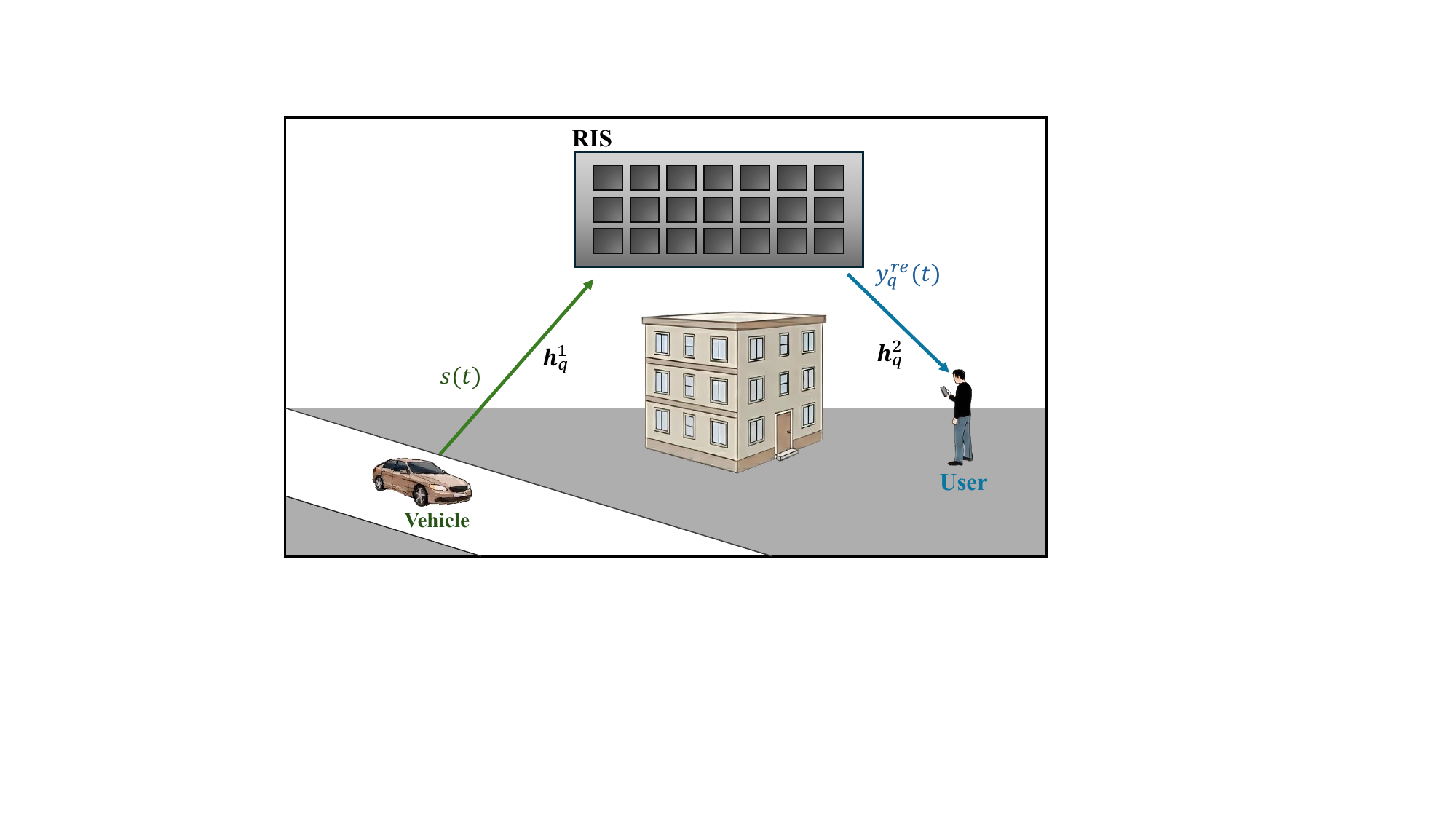}\\
	\caption{System model of the RIS-OTFS-FTN schemes.}
		\label{system model}
\end{figure}

\begin{figure*}
	\centering
	\includegraphics[width=18cm,height=7cm]{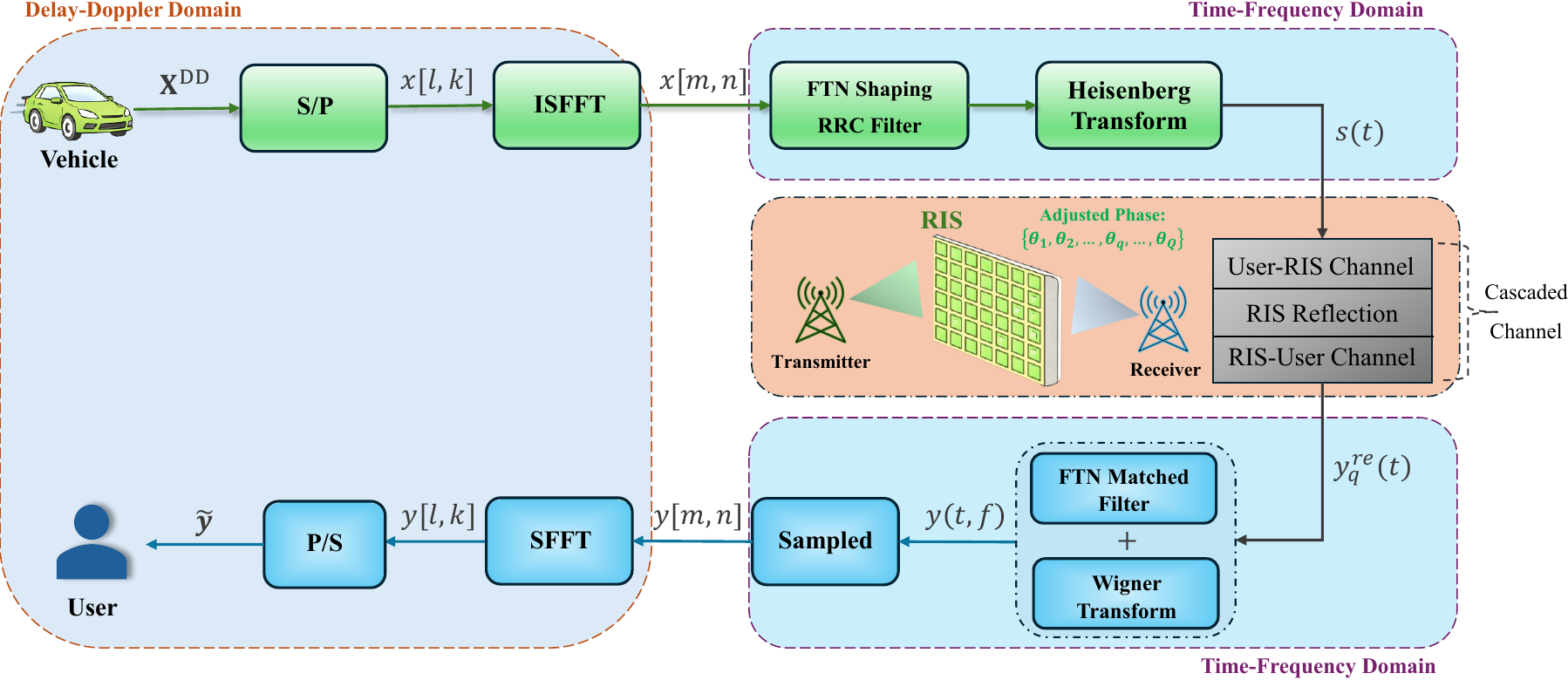}\\
	\caption{Signal processing of the RIS-OTFS-FTN scheme.}
		\label{signal processing}
\end{figure*}

\section{System Model}
\subsection{Transmitted Signals}
For the proposed scheme, we define $M$ and $N$ as the number of delay and  Doppler bins, respectively.
Considering the FTN mapper in our scheme, we also define $\bigtriangleup f$ as the subcarrier spacing and  FTN-specific symbol interval $T_F=T/M=\alpha T_0\,\,( 0 <\mathrm{\alpha} \le 1 )$, where $T_F$ is shorter than that defined under ISI-free Nyquist criterion $T_0$, $T$ represents the frame duration, and $\alpha$ is the compression factor in the time domain. 
Also, let  $\mathbf{X}^{\mathrm{DD}}\in \mathbb{C} ^{M\times N}$  be the matrix of DD domain symbols in each OTFS frame and $\mathbf{X}^{\mathrm{DD}}=\left\{ x\left[ l,k \right] \right\} _{l=0,k=0}^{M-1,N-1}$.
Each $x\left[ l,k \right]$ is modulated by $M$-ary quadrature amplitude modulation (QAM) with average power $E_x=E\left[ \left| x\left[ l,k \right] \right|^2 \right] =\sigma _{x}^{2}$, and placed in the right places after adding cyclic prefix (CP).
By using the inverse symplectic fast Fourier transform (ISFFT), $\mathbf{X}^{\mathrm{DD}}$ can be transformed into the TF domain as
\begin{align}\label{ISFFT}
	X[m,n]=\frac{1}{\sqrt{MN}}\sum_{l=0}^{M-1}{\sum_{k=0}^{N-1}{x[l,k]e^{j2\pi \left( \frac{ml}{M}-\frac{nk}{N} \right)}},}
\end{align}
where $m=0, 1,...,M-1$ and $n=0,1,...,N-1$ are $m$-th and $n$-th time and frequency bin in the TF grid, respectively.
Next, to achieve FTN processing, Heisenberg transform and a root raised-cosine (RRC) pulse shaping filter with a roll-off factor $\beta$ is employed in $X[m,n]$, which is given as
\begin{align} \label{Heisenberg transform}
	s(t)=\sum_{m=0}^{M-1}{\sum_{n=0}^{N-1}X[m,n]g_{tx}(t-n\alpha T_0)e^{j2\pi m\Delta f_0(t-n\alpha T_0)}},
\end{align}
where $g_{tx}(t-n\alpha T_0)$ is the band-limited impulse response shaped by the RRC filter, as the first specific characteristic in our scheme and differing with the rectangular pulses adopted in the conventional schemes.

\subsection{Channel Models}
According to \figref{system model}, an urban environment with many buildings is considered, where we focus on the V2X communications from the vehicle to the fast-moving user and the direct link from vehicle to user is blocked by barriers.
Thus, we consider the time-varying channels from vehicle to the $q$-th reflection element of RIS and from the $q$-th reflection element to user, which are respectively given in DD domain as 
\begin{align} \label{h1 DD}
	\boldsymbol{h}_q^1(\tau _{q}^{1},\nu _{q}^{1})=\sum_{p_1=0}^{P_1-1}{h_{q,p_1}^{1}\delta (\tau _{q}^{1}}-\tau _{q,p_1,}^{1})\delta (\nu _{q}^{1}-\nu _{q,p_1}^{1}),
\end{align}
and
\begin{align} \label{h2 DD}
	\boldsymbol{h}_q^2(\tau _{q}^{2},\nu _{q}^{2})=\sum_{p_2=0}^{P_2-1}{h_{q,p_2}^{2}\delta (\tau _{q}^{2}}-\tau _{q,p_2,}^{2})\delta (\nu _{q}^{2}-\nu _{q,p_2}^{2}),
\end{align}
with $\delta (\cdot)$ representing the Dirac’s delta function, $h_{q,p_1}^1$ and $h_{q,p_2}^2$, $\tau _{q}^{1}$ and $\tau _{q}^{2}$, $\nu _{q}^{1}$ and $\nu _{q}^{2}$  denoting channel gain, delay , and Doppler of $q$-th reflection element for $p_1$-th and $p_2$-th paths,  $P_1$ and $P_2$ representing number of paths for both links.
Moreover, we can further express $\tau _{q}^{1}$ and $\tau _{q}^{2}$, $\nu _{q}^{1}$ and $\nu _{q}^{2}$ as
\begin{align}  \label{delay Doppler}
	\tau _{q,p_1}^{1}=\frac{l_{p_1}}{M\Delta f}\quad \text{and} \quad \nu _{q,p_1}^{1}=\frac{r_{p_1}+\kappa _{p_1}}{NT_F}, \nonumber
	\\
	\tau _{q,p_2}^{2}=\frac{l_{p_2}}{M\Delta f} \quad \text{and} \quad \nu _{q,p_2}^{2}=\frac{r_{p_2}+\kappa _{p_2}}{NT_F},
\end{align}
where $l_{p_1}$ and $l_{p_2}$ represent the delay taps for $p_1$ and $p_2$ paths, $r_{p_1}+\kappa _{p_1}$ and $r_{p_2}+\kappa _{p_2}$ are the Doppler shift taps of $p_1$ and $p_2$ paths.
Note that $l_{p_1}$, $l_{p_2}$, $r_{p_1}$, and $r_{p_2}$ are all integers, but $\kappa _{p_1}$ and $\kappa _{p_2}$ are the fractional parts with the range in $(-\frac{1}{2},\frac{1}{2})$ .
\footnote{In practical wireless channels, Doppler shifts are continuous values that rarely align exactly with the discrete Doppler bins defined by the system. Therefore, a fractional Doppler component is introduced to accurately model the true Doppler shift and avoid performance degradation caused by quantization errors.}
Based on \eqref{h1 DD} and \eqref{h2 DD}, by applying inverse Fourier transform, we can obtain channel responses of vehicle-RIS and RIS-user links in time domain as
\begin{align} \label{h1 T1}
	\boldsymbol{h}_{q}^{1}(t)&=\int_{-\infty}^{\infty}{\boldsymbol{h}_{q}^{1}(\tau _{q}^{1},\nu _{q}^{1})e^{j2\pi \nu _{q}^{1}\left( t-\tau _{q}^{1} \right)}d\nu _{q}^{1}}  \nonumber
	\\
	&=\sum_{p_1=0}^{P_1-1}{h_{q,p_1}^{1}e^{j2\pi \nu _{q,p_1}^{1}\left( t-\tau _{q,p_1}^{1} \right)}\delta (\tau}-\tau _{q,p_1}^{1}),
\end{align}
and 
\begin{align} \label{h1 T2}
	\boldsymbol{h}_{q}^{2}(t)&=\int_{-\infty}^{\infty}{\boldsymbol{h}_{q}^{2}(\tau _{q}^{2},\nu _{q}^{2})e^{j2\pi \nu _{q}^{2}\left( t-\tau _{q}^{2} \right)}d\nu _{q}^{2}}    \nonumber
	\\
	&=\sum_{p_2=0}^{P_2-1}{h_{q,p_2}^{2}e^{j2\pi \nu _{q,p_2}^{2}\left( t-\tau _{q,p_2}^{2} \right)}\delta (\tau}-\tau _{q,p_2}^{2}).
\end{align}

\subsection{Received Signals}
Firstly, let us focus on the received signals on the RIS.
For each reflection element, we define $y_{q}^{ris}(t)$ as the received signal on the $q$-th reflection element corresponding to the incident signal $s(t)$ from vehicle.
Specifically, $y_{l}^{ris}(t)$ in time domain can be expressed as
\begin{align} \label{RIS signals}
	y_{q}^{ris}(t)=\sum_{p_1=\boldsymbol{0}}^{P_1-1}{h_{q,p_1}^{1}e^{j2\pi \nu _{q,p_1}^{1}(t-\tau _{q,p_1}^{1})}s(t}-\tau _{q,p_1}^{1}).
\end{align}
Therefore, after adjusting by the RIS, the reflecting signals to receiver can be expressed as
\begin{align} \label{reflecting signals}
	y_q^{re}(t)&=\beta _qe^{j\theta _q}y_{q}^{ris}(t)   \nonumber
	\\
	&=\beta _qe^{j\theta _q}\sum_{p_1=0}^{P_1-1}{h_{q,p_1}^{1}e^{j2\pi \nu _{q,p_1}^{1}(t-\tau _{q,p_1}^{1})}s(t}-\tau _{q,p_1}^{1}),
\end{align}
where $\beta _qe^{j\theta _q}$ is the reflection coefficient of $q$-th reflection element with amplitude attenuation $\beta _q\in \left[ 0,1 \right]$ and adjusted phase $\theta _q\in \left[ 0,2\pi \right]$.
Next, we can simply obtain the received signals from $q$-th to receiver, which is given as
\begin{align} \label{received signals}
	y_q(t)=&\sum_{p_1=0}^{P_1-1}{\sum_{p_{\mathbf{2}}=0}^{P_2-1}{h_{q,p_2}^{2}e^{j2\pi \nu _{q,p_2}^{2}(t-\tau _{q,p_1}^{1}-\tau _{q,p_2}^{2})}}}\beta _qe^{j\theta _q}  \nonumber
	\\
	&h_{q,p_1}^{1}e^{j2\pi \nu _{q,p_1}^{1}(t-\tau _{q,p_1}^{1}-\tau _{q,p_2}^{2})}s\left( t-\tau _{q,p_1}^{1}-\tau _{q,p_2}^{2} \right)
\end{align}
For convenience, we can further rewrite \eqref{received signals} as
\begin{align} \label{simple received signals}
	&y_q(t) \nonumber
	\\
	&=\Phi _q\sum_{p_1=0}^{P_1-1}{\sum_{p_2=0}^{P_2-1}{h_{q,p_{1,2}}^{\mathrm{eff}}}}e^{j2\pi \nu _{q,p_{1,2}}^{\mathrm{eff}}(t-\tau _{q,p_{1,2}}^{\mathrm{eff}})}s\left( t-\tau _{q,p_{1,2}}^{\mathrm{eff}} \right),
\end{align}
where $\Phi _q=\beta _qe^{j\theta _q}$, $h_{q,p_{1,2}}^{\mathrm{eff}}=h_{q,p_2}^{1}h_{q,p_2}^{2}$, $\nu _{q,p_{1,2}}^{\mathrm{eff}}=\nu _{q,p_1}^{1}+\nu _{q,p_2}^{2}$, and $\tau _{q,p_{1,2}}^{\mathrm{eff}}=\tau _{q,p_1}^{1}+\tau _{q,p_2}^{2}$.
Eventually, by defining $Q$ as number of total reflection elements, the received signals from RIS with total reflection elements to receiver can be given as
\begin{align} \label{total received signals}
	&y(t)=\sum_{q=1}^Q{y_q(t)}  \nonumber
	\\
	&=\sum_{q=1}^Q{\sum_{p_1=0}^{P_1-1}{\sum_{p_{\mathbf{2}}=0}^{P_2-1}{h_{q,p_{1,2}}^{\mathrm{eff}}}}\Phi _qe^{j2\pi \nu _{q,p_{1,2}}^{\mathrm{eff}}(t-\tau _{q,p_{1,2}}^{\mathrm{eff}})}s\left( t-\tau _{q,p_{1,2}}^{\mathrm{eff}} \right)}.
\end{align}
\par
Based on FTN processing at receiver, we consider the window function of RRC filter as $g_{rx}\left( t \right) =g_{tx}^{*}\left( -t \right)$, which is used to  facilitate the conversion of the incoming signal $y(t)$ into TF domain by Wigner transform.
Specifically, it can be expressed as
\begin{align} \label{Wigner}
	y\left( t,f \right) =\int{g_{rx}^{*}\left( t^{\prime}-t \right) y\left( t^{\prime} \right) e^{-j2\pi f\left( t^{\prime}-t \right)}dt^{\prime}}.
\end{align}
Furthermore,  via matched filter at frequency $f=m\Delta f_0$ and time $t=n\alpha T_0$ , signals in \eqref{Wigner} need to be sampled in TF domain, which is given as
\begin{align} \label{Sample}
	y\left[ m,n \right] =y\left( t,f \right) |_{t=n\alpha T_0,f=m\Delta f}.
\end{align}
\par
Finally, by employing symplectic fast Fourier transform (SFFT), we can obtain received signals in DD domain, which is given as
\begin{align} \label{DD received signals}
	y[l,k]=\frac{1}{\sqrt{MN}}\sum_{m=0}^{M-1}{\sum_{n=0}^{N-1}{y[m,n]e^{-j2\pi \left( \frac{ml}{M}-\frac{nk}{N} \right)}}}.
\end{align}
Detail illustration of signal processing in our scheme is shown in \figref{signal processing}.

\subsection{Input-Output Relation in DD Domain}
In order to represent the RIS-assisted OTFS system in the discrete DD domain, we firstly quantize the continuous-time delay $\tau_{q,p_{1,2}}$ and Doppler $\nu_{q,p_{1,2}}$ values to their corresponding discrete taps. 
Given an OTFS grid with $M$ delay bins and $N$ Doppler bins, the resolutions are defined as
\begin{align} \label{delay Doppler bins}
\Delta \tau = \frac{1}{M \Delta f},  \quad  \Delta \nu = \frac{1}{N T_F},
\end{align}
where $\Delta f$ is the subcarrier spacing and $T_F$ is the symbol duration via FTN.
Moreover, for a given path with continuous delay and Doppler values $(\tau_{q,p_{1,2}}, \nu_{q,p_{1,2}})$, the corresponding discrete delay tap $\varepsilon _{q,p_{1,2}}$ and Doppler tap $k_{q,p_{1,2}}$ are given by
\begin{align} \label{delay Doppler taps}
\varepsilon _{q,p_{1,2}} = \left\lfloor \tau_{q,p_{1,2}} M \Delta f \right\rfloor, \quad k_{q,p_{1,2}} = \left\lfloor \nu_{q,p_{1,2}} N T_F \right\rfloor.
\end{align}
\par
After this, we continue to formulate the input-output relationship in the discrete DD domain. 
Considering the cascaded vehicle-RIS-user channels, the received time-domain signal at $a$-th element in DD 2D grid can be expressed as \equref{received signal at DD element} at the top in next page,
\begin{figure*}[ht] 
	\centering 
	\vspace*{0pt} 
	\begin{align} \label{received signal at DD element}
		y[a]=\sum_{q=1}^Q{\sum_{p_1=0}^{P_1-1}{\sum_{p_2=0}^{P_2-1}{h_{q,p_{1,2}}^{\mathrm{eff}}\Phi _q\,e^{j2\pi \frac{k_{q,p_{1,2}}(a-\varepsilon _{q,p_{1,2}})}{NM}}g\left( t-\left( a+\varepsilon _{q,p_{1,2}}  \right) T_{\mathrm{f}} \right) s\left[ [a-\varepsilon _{q,p_{1,2}}]_{NM} \right]}},}
	\end{align}
\end{figure*}
where $s[a]$ is the time-domain symbol at $a$-th element in DD 2D grid. 
Moreover, \eqref{received signal at DD element} can be compactly written in vector form as
\begin{align} \label{vectorized received signals}
	\tilde{\mathbf{y}}=\sum_{q=1}^Q{\Phi _q\tilde{\mathbf{H}}_q\tilde{\mathbf{s}}},
\end{align}
where $\tilde{\mathbf{s}}\in \mathbb{C} ^{NM\times 1}$ and $\tilde{\mathbf{y}}\in \mathbb{C} ^{NM\times 1}$  respectively represents the transmitted and the received data vector in time domain with $\tilde{\mathbf{y}}=\left\{ y\left[ a \right] \right\} _{a=0}^{MN-1}$ and $\tilde{\mathbf{s}}=\left\{ s\left[ a \right] \right\} _{a=0}^{MN-1}$, and $\tilde{\mathbf{H}}_q\in \mathbb{C} ^{NM\times NM}$ is the cascaded channel matrix with cascaded coefficient $h_q$.
\par
Applying OTFS modulation and demodulation, e.g., ISFFT in \equref{ISFFT}, Heisenberg transform in \equref{Heisenberg transform}, Wigner transform with FTN-included ISI in \equref{Wigner} and sampling, we can obtain cascaded channel in TF domain as $\mathbf{H}^{\mathrm{eff}}_{\mathrm{TF}}$ with $\left\{ \mathbf{H}_{\mathrm{TF}}^{\mathrm{eff}} \right\} _{\boldsymbol{n},\boldsymbol{m}}=\mathbf{H}_q[m,n]$.
Given that DD-domain channel gain matrix $\mathbf{H}_{\mathrm{TF},q} $ for the $q $-th RIS element, in which non-zero entry corresponds to the complex gain, denoted as $h_{q,p_{1,2},a}$, which is expressed as \equref{element of channel coefficiency} at the top of next page,
\begin{figure*}[ht] 
	\centering 
	\vspace*{0pt} 
	\begin{align} \label{element of channel coefficiency}
		\mathbf{H}_{\mathrm{TF},q}[m,n] = \left\{
		\begin{aligned}
			&\sum_{p_1=0}^{P_1-1}\sum_{p_2=0}^{P_2-1}
			h_{q,p_{1,2},a}^{\mathrm{eff}}
			e^{j2\pi \frac{k_{q,p_{1,2}}(a-\varepsilon _{q,p_{1,2}})}{NM}}
			g\!\left(mT_{\mathrm{f}}-(n+\varepsilon _{q,p_{1,2}})T_{\mathrm{f}}\right), \,
			a=[n-m]_{NM},\ \zeta \ge \acute{\zeta} \\[8pt]
			&\sum_{p_1=0}^{P_1-1}\sum_{p_2=0}^{P_2-1}
			h_{q,p_{1,2},a}^{\mathrm{eff}}
			e^{j2\pi \frac{k_{q,p_{1,2}}(a-\varepsilon _{q,p_{1,2}})}{NM}}
			g\!\left(mT_{\mathrm{f}}-(n+\varepsilon _{q,p_{1,2}})T_{\mathrm{f}}\right), \,
			a=[n-m+M]_{NM},\ \zeta < \acute{\zeta} \\[8pt]
			&0,\quad \mathrm{otherwise}
		\end{aligned}
		\right.
	\end{align}
	\hrulefill 
\end{figure*}
with $q=0,1,...,Q$,  $n=0,1,...,NM-1$, $m=0,1,...,MN-1$, $\zeta=n-M\lfloor \frac{n}{M} \rfloor $, $\zeta^{\prime}=m-M\lfloor \frac{m}{M} \rfloor$, and $g(t)\triangleq g_{\mathrm{tx}}(t)\star g_{\mathrm{rx}}^{*}(-t)$.
\par
Let $\mathbf{X}^{\mathrm{DD}} \in \mathbb{C}^{M \times N}$ denote the transmitted symbol matrix in the DD domain, whose $(m,n)$-th entry carries the data symbol mapped onto the DD grid.
For notational convenience, we define the vectorized DD-domain symbol block as $\mathbf{x}^{\mathrm{DD}} \triangleq \mathrm{vec}\!\left(\mathbf{X}^{\mathrm{DD}}\right) \in \mathbb{C}^{MN \times 1}$.
Next, applying SFFT as \equref{DD received signals}, the matrix-form system output in the DD domain becomes \cite{ref27}:
\begin{align} \label{DD output}
	\mathbf{y} = \mathbf{H}^{\mathrm{eff}}\,\mathbf{x}^{\mathrm{DD}} + \mathbf{z}^{\mathrm{DD}},
\end{align}
where $\mathbf{H}^{\text{eff}} =  [\mathbf{H}_1, \mathbf{H}_2, \dots, \mathbf{H}_Q] \, \boldsymbol{\Phi}$ with $\mathbf{\Phi }=\mathrm{diag}\left( \beta _1e^{j\theta _1},\dots ,\beta _Qe^{j\theta _Q} \right) \otimes \mathbb{I} _{NM}$ and $\mathbf{H}_q \in \mathbb{C} ^{NM\times NM}$ is the channel  coefficient matrices of the cascaded channel 
\footnote{Specifically, each element of the effective channel matrix $\mathbf{H}_q \in \mathbb{C}^{NM \times NM}$ at row $u$ and column $v$ (where $0 \le u, v \le NM-1$) is explicitly given by the discretized ambiguity function as $[\mathbf{H}_q]_{u,v} = \sum_{p=1}^{L} h_p e^{j \vartheta_p} \mathcal{A}_{g}\left( \frac{u-v}{NM}T_f - \tau_p, \frac{u-v}{NM}\Delta f - \nu_p \right)$,
where $L$ is the number of propagation paths, $h_p$ is the path gain, $\vartheta_p$ includes the RIS phase shift, $(\tau_p, \nu_p)$ are the delay and Doppler shifts of the $p$-th path and $\mathcal{A}_{g}(\tau, \nu) = \int g_{\text{tx}}(t) g_{\text{rx}}^*(t-\tau) e^{-j 2\pi \nu t} dt$ represents the cross-ambiguity function of the transmit and receive pulse shaping filters, which inherently accounts for the inter-symbol interference induced by FTN signaling.}, 
and 
$\mathbf{z}_{\mathrm{DD}}=\left( \mathbf{F}_N\otimes \mathbf{F}_{M}^{H} \right) \mathbf{z}_{\mathrm{TF}}$, 
$\mathbf{z}_{\mathrm{TF}}\in \mathbb{C} ^{MN\times 1}$ is the vector of time domain additive colored complex Gaussian noise with covariance $\mathbb{E}[\mathbf{z}_{\mathrm{TF}}\mathbf{z}_{\mathrm{TF}}^{H}] = \sigma^2 \mathbf{G}$ and $\mathbf{G}$ denotes the Toeplitz correlation matrix determined by the FTN pulse shaping.
Moreover, $\mathbf{F}_N \in \mathbb{C}^{N \times N}$ and $\mathbf{F}_M \in \mathbb{C}^{M \times M}$ denote the normalized Discrete Fourier Transform (DFT) matrices. 
The $(k, l)$-th entry of $\mathbf{F}_N$ is given by	$[\mathbf{F}_N]_{k,l} = \frac{1}{\sqrt{N}} e^{-j \frac{2\pi}{N}(k-1)(l-1)}, \quad 1 \le k, l \le N$,
and $\mathbf{F}_M$ is defined similarly with dimension $M$. 
\par
We note that FTN signaling disrupts the standard OTFS block-circulant property, resulting in a block-Toeplitz effective channel $\mathbf{H}^{\mathrm{eff}}$. 
Although this precludes circular convolution-based equalization, the matrix remains sparse and banded due to localized pulse shaping, which is efficiently handled by the adopted LMMSE detector.

\subsection{Linear Minimum Mean Square Error Detector}
In this work, taking into account the trade-off between complexity and error performance, the Linear Minimum Mean Square Error (LMMSE) detector is employed for the proposed scheme, instead of the maximum likelihood (ML) or message-passing (MP) detectors. 
Although the vehicle-RIS channel matrix $\mathbf{H}_1$ and RIS-user channel matrix $\mathbf{H}_2$ are individually sparse, the cascaded effective channel matrix $\mathbf{H}^{\mathrm{eff}} = \mathbf{H}_2 \mathbf{\Phi} \mathbf{H}_1$ loses its sparsity after matrix multiplication \cite{ref23}. 
This leads to increased complexity and degraded error performance for the MP detector, as the density of the factor graph increases significantly.
Second, the optimal ML detector requires an exhaustive search over all potential symbol combinations, calculating Euclidean distances for the entire $NM$-dimensional vector space. 
This results in prohibitively high computational complexity that scales exponentially with the grid size, rendering it impractical for high-mobility OTFS systems.
Consequently, we adopt the LMMSE detector \cite{ref31}, which minimizes the mean square error under the constraint of a linear filter. 
The LMMSE detection based on the structured colored noise is given by:
\begin{align} \label{LMMSE}
	\hat{\mathbf{x}}^{\mathrm{DD}} = \left( \mathbf{H}_{\mathrm{eff}}^{\dagger} \mathbf{H}_{\mathrm{eff}} + \sigma^2 \mathbf{G} \right)^{-1} \mathbf{H}_{\mathrm{eff}}^{\dagger} \mathbf{y},
\end{align}
where $\hat{\mathbf{x}}^{\mathrm{DD}} \in \mathbb{C}^{NM \times 1}$ denotes estimated signal vector. 
$\mathbf{H}_{\mathrm{eff}}^{\dagger}$ is the conjugate transpose of the effective channel matrix $\mathbf{H}_{\mathrm{eff}}$.

\section{Performance Analysis}
In this section, the analytical performance of AFER and spectral efficiency is provided with closed-form expressions.
Moreover, the CCDF of the PAPR and IBO  are also given respectively  in sub-section \textit{C } and \textit{D} of this section.
More interesting findings and insights are also given in this section.

\subsection{AFER}
While the ML detector is challenging to implement in both practical systems and simulations, its theoretical performance can still be characterized by deriving the upper bound of the AFER, which allows us to gain valuable insights into its fundamental behavior.
\footnote{Our analysis uses the optimal ML detector as a benchmark. The derived ML upper bound rigorously captures the system's inherent diversity order and coding gain, defining its performance potential. The LMMSE simulations, meanwhile, demonstrate the practical performance achievable under specific receiver constraints.}
We define $E_f$ as the energy per frame with $E_f=MNE_x$.
Let $\mathbf{x}^{\mathrm{DD}} = \mathrm{vec}(\mathbf{X}^{\mathrm{DD}}) \in \mathbb{C}^{NM \times 1}$ denote the vectorized transmitted symbol block. 
According to the system model, $\mathbf{x}^{\mathrm{DD}}$ is drawn from a finite codebook $\mathcal{S}$ with size $|\mathcal{S}|$, where $\mathcal{S} =\{\mathbf{x}_{1}^{\mathrm{DD}},\mathbf{x}_{2}^{\mathrm{DD}},...,\mathbf{x}_{|\mathcal{S} |}^{\mathrm{DD}}\}$.
\par
Given that the FTN-induced noise $\mathbf{z}$ follows a colored Gaussian distribution with covariance $\mathbf{\Sigma}_z$, the optimal ML detector minimizes the Mahalanobis distance rather than the Euclidean distance. Thus, the detection rule is expressed as
\begin{align} \label{ML detector}
	\hat{\mathbf{x}}^{\mathrm{DD}}
	&=\mathrm{arg}\min_{\mathbf{x}^{\mathrm{DD}}\in \mathcal{S}} (\mathbf{y}-\mathbf{H}^{\mathrm{eff}}\mathbf{x})^H\mathbf{\Sigma }_{z}^{-1}(\mathbf{y}-\mathbf{H}^{\mathrm{eff}}\mathbf{x})    \nonumber
	\\
	&=\mathrm{arg}\min_{\mathbf{x}^{\mathrm{DD}}\in \mathcal{S}} \left\| \mathbf{y}-\mathbf{H}^{\mathrm{eff}}\mathbf{x}^{\mathrm{DD}} \right\| _{\mathbf{\Sigma }_{z}^{-1}}^{2},
\end{align}
where $\| \mathbf{u} \|_{\mathbf{\Sigma}_z^{-1}}^2 \triangleq \mathbf{u}^{\dagger} \mathbf{\Sigma}_z^{-1} \mathbf{u}$ denotes the squared Mahalanobis norm.
Based on the classical union bound technique, the AFER is upper bounded as
\begin{equation} \label{FER union_bound}
	P_f \le \frac{1}{|\mathcal{S} |}\sum_i{\sum_{j\ne i}{P(\mathbf{x}_{i}^{\mathrm{DD}}\rightarrow \mathbf{x}_{j}^{\mathrm{DD}})}},
\end{equation}
where $P(\mathbf{x}_{i}^{\mathrm{DD}}\rightarrow \mathbf{x}_{j}^{\mathrm{DD}})$ is the pairwise error probability (PEP).
Substituting $\mathbf{y} = \mathbf{H}^{\mathrm{eff}}\mathbf{x}^{\mathrm{DD}}_i + \mathbf{z}$ into \eqref{ML detector}, the PEP event corresponds to the inequality:
\begin{equation}
	\left\| \mathbf{H}^{\mathrm{eff}}(\mathbf{X}^{\mathrm{DD}}_i - \mathbf{X}^{\mathrm{DD}}_j) + \mathbf{z} \right\|_{\mathbf{\Sigma}_z^{-1}}^2 
	\le \left\| \mathbf{z} \right\|_{\mathbf{\Sigma}_z^{-1}}^2.
\end{equation}
Expanding the Mahalanobis norm and canceling the common term $\mathbf{z}^{\dagger}\mathbf{\Sigma}_z^{-1}\mathbf{z}$ yields
\begin{equation}
	2\Re \left\{ \mathbf{z}^{\dagger}\mathbf{\Sigma}_z^{-1} \mathbf{H}^{\mathrm{eff}}\Delta_{ij} \right\} \le - \Delta_{ij}^{\dagger} (\mathbf{H}^{\mathrm{eff}})^{\dagger} \mathbf{\Sigma}_z^{-1} \mathbf{H}^{\mathrm{eff}} \Delta_{ij},
\end{equation}
where $\Delta_{ij} = \mathbf{x}^{\mathrm{DD}}_i - \mathbf{x}^{\mathrm{DD}}_j$ is the error difference vector.
The term on the left-hand side, denoted as $w = \mathbf{z}^{\dagger}\mathbf{\Sigma}_z^{-1} \mathbf{H}^{\mathrm{eff}}\Delta_{ij}$, is a scalar Gaussian random variable with zero mean. Its variance is explicitly derived as
\begin{align}
	\sigma_w^2 &= \mathbb{E}[w w^{\dagger}] \notag \\
	&= \Delta_{ij}^{\dagger} (\mathbf{H}^{\mathrm{eff}})^{\dagger} \mathbf{\Sigma}_z^{-1} \underbrace{\mathbb{E}[\mathbf{z}\mathbf{z}^{\dagger}]}_{\mathbf{\Sigma}_z} \mathbf{\Sigma}_z^{-1} \mathbf{H}^{\mathrm{eff}} \Delta_{ij} \notag \\
	&= \Delta_{ij}^{\dagger} (\mathbf{H}^{\mathrm{eff}})^{\dagger} \mathbf{\Sigma}_z^{-1} \mathbf{H}^{\mathrm{eff}} \Delta_{ij}.
\end{align}
Using the Q-function and Chernoff bound \cite{ref32, ref33}, the PEP is finally upper-bounded by
\begin{equation} \label{PEP bound}
	\begin{aligned}
		P(\mathbf{x}_{i}^{\mathrm{DD}}\rightarrow \mathbf{x}_{j}^{\mathrm{DD}})&=Q\left( \sqrt{\frac{\Delta_{ij}^{\dagger} (\mathbf{H}^{\mathrm{eff}})^{\dagger} \mathbf{\Sigma}_z^{-1} \mathbf{H}^{\mathrm{eff}} \Delta_{ij}}{2}} \right) 
		\\
		&\le \exp \left( -\frac{\Delta_{ij}^{\dagger} (\mathbf{H}^{\mathrm{eff}})^{\dagger} \mathbf{\Sigma}_z^{-1} \mathbf{H}^{\mathrm{eff}} \Delta_{ij}}{4} \right) .
	\end{aligned}
\end{equation} 
In the other hand, let us define the frame difference vector as $\Delta_{ij} = \mathbf{X}_{i}^{\mathrm{DD}} - \mathbf{X}_{j}^{\mathrm{DD}} $.
Then the exponent term in \eqref{PEP bound} can be rewritten as a quadratic form
\begin{equation}\label{quad_form_vec}
	\Delta_{ij}^\dagger \mathbf{H}_{\mathrm{eff}}^\dagger \mathbf{\Sigma}_z^{-1} \mathbf{H}_{\mathrm{eff}} \Delta_{ij}
	= \operatorname{vec}(\mathbf{H}_{\mathrm{eff}})^\dagger 
	\bigl(\Delta_{ij}\Delta_{ij}^\dagger \otimes \mathbf{\Sigma}_z^{-1}\bigr)
	\operatorname{vec}(\mathbf{H}_{\mathrm{eff}}).
\end{equation}
Deriving an exact AFER is mathematically intractable since the single-element cascaded channel follows a correlated double-Rayleigh distribution. However, invoked by the Central Limit Theorem (CLT), the aggregated response from $Q$ independent RIS elements asymptotically converges to a Gaussian distribution. Thus, we can safely approximate the non-zero entries in $\mathbf{H}_{\mathrm{eff}}$ as $\tilde{h}_n \sim \mathcal{CN}(0, \sigma_h^2)$.
Then, we can use the well-known moment-generating function (MGF) of complex Gaussian quadratic forms. 
For any Hermitian positive semidefinite matrix $\mathbf{A}$, we have
\begin{equation}  \label{eq: A 1}
	\mathbb{E}\!\left[e^{-\operatorname{vec}(\mathbf{H}_{\mathrm{eff}})^\dagger \mathbf{A} \operatorname{vec}(\mathbf{H}_{\mathrm{eff}})}\right]
	= \det\!\big(\mathbf{I}+\sigma_h^2 \mathbf{A}\big)^{-1}.
\end{equation}
Applying this result with $\mathbf{A}=\tfrac{1}{4}\bigl(\Delta_{ij}\Delta_{ij}^\dagger \otimes \mathbf{\Sigma}_z^{-1}\bigr)$, the PEP expression can be further expressed as
\begin{equation}\label{PEP_avg_det}
	P(\mathbf{X}_{i}^{\mathrm{DD}} \!\to\! \mathbf{X}_{j}^{\mathrm{DD}})
	\le \det\!\left(\mathbf{I}+\tfrac{\sigma_h^2}{4}\,\Delta_{ij}\Delta_{ij}^\dagger \otimes \mathbf{\Sigma}_z^{-1}\right)^{-1}.
\end{equation}
\par
To obtain a more interpretable expression, let 
$\{\lambda_n^{(ij)}\}_{n=1}^{r_{ij}}$ denote the non-zero eigenvalues of 
$\Delta_{ij}\Delta_{ij}^\dagger$, and let 
$\{\nu_m\}_{m=1}^{MN}$ denote the eigenvalues of $\mathbf{\Sigma}_z^{-1}$. 
Using the Kronecker product property in $\mathrm{spec}\left( A\otimes B \right) =\{\lambda _n(A)\nu _m(B)\}$, \eqref{PEP_avg_det} can be equivalently expressed as
\begin{equation}\label{PEP_prod}
	P(\mathbf{X}_{i}^{\mathrm{DD}} \!\to\! \mathbf{X}_{j}^{\mathrm{DD}})
	\le \prod_{n=1}^{r_{ij}}\prod_{m=1}^{MN}\left(1+\tfrac{\sigma_h^2}{4}\,\lambda_n^{(ij)}\,\nu_m\right)^{-1}.
\end{equation}
\par
Finally, substituting \eqref{PEP_prod} into the union bound \eqref{FER union_bound}, 
we obtain the AFER upper bound as
\begin{equation}\label{AFER_final}
	P_f \le \frac{1}{|\mathcal{S}|}\sum_{i=1}^{|\mathcal{S}|}\sum_{\substack{j=1\\ j\ne i}}^{|\mathcal{S}|}
	\prod_{n=1}^{r_{ij}}\prod_{m=1}^{MN}\left(1+\tfrac{\sigma_h^2}{4}\,\lambda_n^{(ij)}\,\nu_m\right)^{-1}.
\end{equation}
\remark{A higher $E_f$ enhances signal strength, reducing the error probability, and this effect is indirectly reflected through the channel gain and symbol energy distribution.
	Besides, due to $ E_f = MN E_x $, when \( M \) and \( N \) increase, $ E_f $ grows linearly, and \( \text{SNR} = \frac{E_f}{\sigma^2} \propto MN \) increases. 
	Note that $\mathbf{\Sigma}_z$ and $\mathbf{G}$ are positive-definite matrices.
	In the formula, the number of terms in $\prod_{m=1}^{MN} $ rises from $ MN $ to $ M'N' $ with more $ \nu_m $ terms involved, leading to lower AFER.  }
\remark{From the eigenvalue decomposition of the PEP upper bound in \eqref{PEP bound}, let $r_{ij} \triangleq \mathrm{rank}(\boldsymbol{\Delta}_{ij}\boldsymbol{\Delta}_{ij}^\dagger) = \mathrm{rank}(\boldsymbol{\Delta}_{ij})$ denote the number of non-zero eigenvalues of the error difference matrix $\boldsymbol{\Delta}_{ij}\boldsymbol{\Delta}_{ij}^\dagger$.
	Given that the FTN-induced noise covariance $\mathbf{\Sigma}_z = \sigma^2 \mathbf{G}(\alpha)$ is positive definite, its inverse eigenvalues scale as $\nu_m(\alpha) = \mu_m(\alpha)/\sigma^2$, where $\{\mu_m(\alpha)\}$ are the eigenvalues of $\mathbf{G}^{-1}(\alpha)$ which are independent of the noise power $\sigma^2$.
	In the high-SNR regime, each term in the product expansion behaves as $(1 + c \lambda_n^{(ij)} \nu_m(\alpha))^{-1} \doteq \mathrm{SNR}^{-1}$. This yields the asymptotic decay rate of the error probability as
	\begin{equation}
			P\!\left(\mathbf{x}^{\mathrm{DD}}_i \!\to\! \mathbf{x}^{\mathrm{DD}}_j\right) \doteq \mathrm{SNR}^{-MN \cdot r_{ij}}.
	\end{equation}
	Thus, asymptotic diversity order is given by $d = MN \cdot \min_{i \neq j} r_{ij}$.
	Crucially, the FTN acceleration factor $\alpha$ modifies the eigen-spectrum $\{\mu_m(\alpha)\}$ of the correlation matrix $\mathbf{G}(\alpha)$, thereby affecting coding gain. 
	However, it does not alter the diversity exponent (slope) as long as $\mathbf{G}(\alpha)$ remains full rank.}

\begin{figure}
	\centering
	\includegraphics[width=8.5cm,height=7cm]{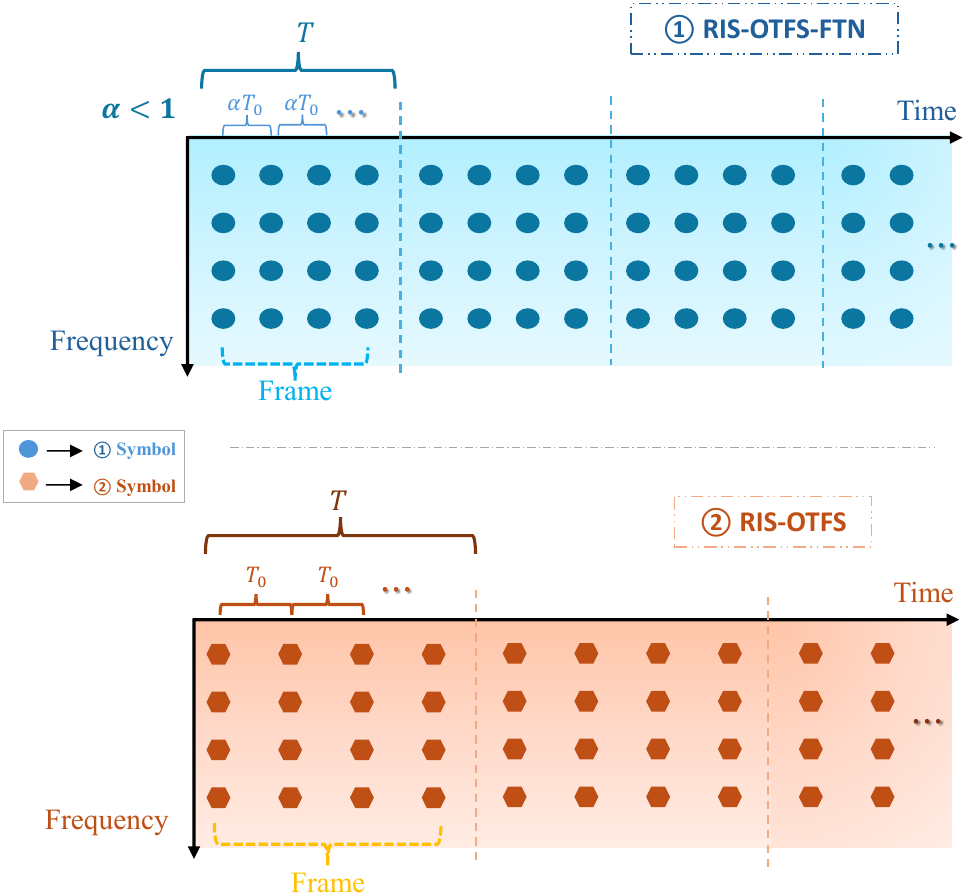}\\
	\caption{Comparison between RIS-OTFS schemes with and without FTN in TF domain.}
	\label{fig:FTN}
\end{figure}

\subsection{Spectral Efficiency}
In order to obtain the final expression of spectral efficiency, let us firstly derive the expression of mutual information for per frame of our scheme.
According to \eqref{DD output}, the mutual information between $\mathbf{X}_{\mathrm{DD}}$ and $\mathbf{y}$ is defined as
\begin{equation} \label{mutual information 1}
	I(\mathbf{X}_{\mathrm{DD}};\mathbf{y})=h(\mathbf{y})-h(\mathbf{y}\mid \mathbf{X}_{\mathrm{DD}}),
\end{equation}
where $h(\cdot)$ denotes the differential entropy. 
Here, for clear comparisons and presentation, we assume that the input symbols $\mathbf{X}_{\mathrm{DD}}$ are zero-mean circularly symmetric complex Gaussian with covariance matrix $\mathbf{\Sigma }_X$ and the noise vector $\mathbf{z}\sim \mathcal{C} \mathcal{N} (0,\mathbf{\Sigma }_{\mathbf{z}})$ is independent of $\mathbf{X}_{\mathrm{DD}}$. 
Given $\mathbf{X}^{\mathrm{DD}}$, the conditional distribution is $\mathbf{y}\mid \mathbf{X}_{\mathrm{DD}}\sim \mathcal{C} \mathcal{N} \left( \mathbf{H}^{\mathrm{eff}}\mathbf{X}^{\mathrm{DD}},\,\mathbf{\Sigma }_{\mathbf{z}} \right) $, so that we have conditional entropy:
\begin{equation} \label{conditional entropy}
	\begin{aligned}
	h(\mathbf{y}\mid \mathbf{X}^{\mathrm{DD}})&=h(\mathbf{z}_{\mathrm{DD}}) 
	\\
	&=\log _2\left( \left( \pi e \right) ^{MN}\det \Bigl( \mathbf{\Sigma }_{\mathbf{z}} \Bigr) \, \right)  
	\\
	&=NM\log _2\left( \pi e \right) +\log _2\left( \det \Bigl( \mathbf{\Sigma }_{\mathbf{z}}  \Bigr) \right) ,
	\end{aligned}
\end{equation}
where $e$ denotes Euler's number, as the base of natural logarithms with approximate value in 2.7183.
\par
Unconditionally, $\mathbf{y}$ is Gaussian with covariance matrix $\mathbf{H}_{\mathrm{eff}}\,\mathbf{\Sigma }_X\,\mathbf{H}_{\mathrm{eff}}^{H}+\mathbf{\Sigma }_{\mathbf{z}}$.
Thus, we can obtain the marginal entropy $h(\mathbf{y})$ as
\begin{equation} \label{marginal entropy}
	\begin{aligned}
		h(\mathbf{y})&=\log _2\left( \left( \pi e \right) ^{MN}\det \Bigl( \mathbf{H}^{\mathrm{eff}}\mathbf{\Sigma }_X\mathbf{H}_{\mathrm{eff}}^{\dagger}+\mathbf{\Sigma }_{\mathbf{z}} \Bigr) \, \right)
		\\
		&=\log _2\det \Bigl( \mathbf{H}^{\mathrm{eff}}\mathbf{\Sigma }_X\mathbf{H}_{\mathrm{eff}}^{\dagger}+\mathbf{\Sigma }_{\mathbf{z}} \Bigr) +NM\log _2\left( \pi e\, \right) .
	\end{aligned}
\end{equation}
\par
Next, by subtracting the conditional entropy \eqref{conditional entropy} from the marginal entropy \eqref{marginal entropy} yields the final expression of mutual information $\mathrm{I(}\mathbf{X}^{\mathrm{DD}};\mathbf{y})$ as
\begin{equation} \label{mutual information 2}
	\begin{aligned}
		\mathrm{I(}\mathbf{X}^{\mathrm{DD}};\mathbf{y})&=\log _2\det \Bigl( \mathbf{H}^{\mathrm{eff}}\mathbf{\Sigma }_X\mathbf{H}_{\mathrm{eff}}^{\dagger} \Bigr) -\log _2\left( \det \Bigl( \mathbf{\Sigma }_{\mathbf{z}}  \Bigr) \right)
		 \\
		&=\log _2\det \left( \mathbf{I}_{MN}+\mathbf{\Sigma }_{\mathbf{z}_{\mathrm{DD}}}^{-1}\mathbf{H}_{\mathrm{eff}}\mathbf{\Sigma }_X\mathbf{H}_{\mathrm{eff}}^{\dagger} \right)
		\\
		&=\log _2\det \left( \mathbf{I}_{MN}+\frac{1}{\sigma ^2}\mathbf{G}^{-1}\mathbf{H}^{\mathrm{eff}}\mathbf{\Sigma }_X\mathbf{H}_{\mathrm{eff}}^{\dagger} \right).
	\end{aligned}
\end{equation} 
To simplify, perform eigenvalue decomposition (EVD) on $\mathbf{G}=\mathbf{V\Lambda V}^H$, where $\mathbf{V}=\mathrm{diag}\left\{  \lambda^1_0, \lambda^1_1, ... , \lambda^1_MN-1   \right\}$ with $i$-th $\lambda^1_i$ non-zero eigenvalues of $\mathbf{G}$.
Then, given that $\mathbf{B}=\mathbf{\Lambda }^{-1/2}\mathbf{V}^H\mathbf{H}^{\mathrm{eff}}$, \equref{mutual information 2} can be further rewritten as $I(\mathbf{X}_{\mathrm{DD}};\mathbf{y})=\log _2\det \left( \mathbf{I}_{MN}+\frac{E_x}{\sigma ^2}\mathbf{B}^H\mathbf{B} \right)$.
Let $\mathbf{B}^H\mathbf{B}=\mathbf{U\Xi U}^H$ with $\mathbf{\Xi }=\mathrm{diag}\left\{ \mathrm{\xi}_0,\mathrm{\xi}_1,..., \mathrm{\xi}_{MN-1} \right\}$ and $\mathrm{\xi}_i$ being $i$-th non-zeros eigenvalue of $\mathbf{B}^H\mathbf{B}$. 
\equref{mutual information 2} can be simplified as
\begin{equation} \label{finally mutual information}
	\begin{aligned}
		I(\mathbf{X}_{\mathrm{DD}};\mathbf{y})=\sum_{n=0}^{MN-1}{\log _2\left( 1+\frac{E_x\xi _n}{\sigma ^2} \right) }.
	\end{aligned}
\end{equation} 
\par
Eventually, by assuming the energy of each frame is evenly distributed to each symbol, we have a covariance matrix for per frame as $E_x\,=\small{\frac{E_f}{MN}}$, the achievable spectral efficiency of the RIS-OTFS-FTN scheme is defined as the total mutual information per frame normalized by the frame duration as
\begin{align} \label{spectral efficiency}
	\eta =\frac{1}{M\bigtriangleup fN\alpha T_0}\sum_{n=0}^{MN-1}{\log _2\left( 1+\frac{E_f\xi _n}{MN\sigma ^2} \right)}.
\end{align}

\remark{According to \eqref{spectral efficiency}, we can easily find that the spectral efficiency of our scheme increases with the decrease of $\alpha$, which is different from the conventional scheme without FTN due to the absence of $\alpha$. This can also be observed in \figref{fig:FTN}, where we can clearly see that the RIS-OTFS-FTN scheme can transmit more symbols within the same time duration compared to the scheme without FTN, thus enhancing the spectral efficiency. Moreover, it is investigated that the spectral efficiency of our scheme gets much better in the high-SNR regime, where $E_x \gg \sigma ^2$. 
\textcolor{red}{Besides, under a given average transmit power, changing the size of $N$ directly affects the total frame energy $E_f$, e.g., decreasing $N$ makes lower $E_f$. 
Note that details of the expected total energy of an OTFS-FTN frame $E_f$ is provided in Appendix A.
Since $M$ and $N$ play different roles in various terms of \eqref{spectral efficiency}, it is non-trivial to straightforwardly figure out how the spectral efficiency varies with the grid size $M \times N$ at a glance. Detailed simulation results and further discussions on this phenomenon are provided in Sec. \Rmnum{5}.}}

\subsection{PAPR and CCDF}
The PAPR is a critical metric for evaluating the efficiency of power amplifiers in communication systems, particularly in the proposed RIS-OTFS-FTN scheme, where FTN signaling introduces unique signal dynamics \cite{ref34}. 
Attention should be paid to the time-domain transmitted signal $s(t)$, as the PAPR is a power metric computed based on the actual transmission waveform.
The CCDF provides an intuitive measure of the likelihood of encountering large peaks in the transmit waveform, which is critical for evaluating power amplifier back-off requirements. 
Also, it quantifies the probability that the PAPR exceeds a specified threshold, offering insights into the statistical characteristics of the signal \cite{ref34}. 
In this subsection, we derive the analytical framework for computing the CCDF of the PAPR for the proposed scheme, accounting for the effects of FTN compression and RRC pulse shaping.
\par
Based on \equref{ISFFT}, the expression of PAPR is defined as
\begin{equation} \label{PAPR}
	\mathrm{PAPR}=\frac{\max |s(t)|^2}{\mathbb{E} [|s(t)|^2]},
\end{equation}
where $\mathbb{E}[|s(t)|^2]$ represents the average power of the signal. 
For a given threshold $\gamma_0$, the PAPR can be obtained as
\begin{equation} \label{CCDF}
	\begin{aligned}
		\mathrm{CCDF(}\gamma _0) &=\mathrm{Pr(PAPR}>\gamma _0)
		\\
		&=1-\mathrm{Pr}\left( \max_{t\in [0,T]} \frac{|s(t)|^2}{\mathbb{E} [|s(t)|^2]}\le \gamma _0 \right) .
	\end{aligned}
\end{equation}
\par
To obtain the CCDF curve, a large number of frames are generated using Monte Carlo simulations. 
For each frame, the PAPR is calculated, and the empirical distribution is estimated by counting the proportion of frames whose PAPR exceeds a given threshold $\gamma_0$.

\subsection{IBO}
The IBO is a key parameter for assessing the linearity margin of PA and it directly relates to the PAPR of the transmit waveform. 
In this subsection we distinguish two useful IBO definitions used in our analysis: (i) the \emph{minimum required IBO} that guarantees no clipping of waveform peaks, and (ii) the \emph{available IBO} that results from a given PA saturation level and the transmitter average power. 
Based on \equref{PAPR}, the required IBO for linear operation is mainly determined by the PAPR of the transmitted signal plus a design margin to avoid nonlinear distortion, expressed as
\begin{equation}
	\mathrm{IBO}_{\mathrm{req}} = \mathrm{PAPR} + \gamma_\mathrm{margin},
\end{equation}
where $\gamma_\mathrm{margin}$ typically ranges from 1 to 3 dB to accommodate implementation imperfections.
\par
Meanwhile, the available IBO, which is related to the transmitter hardware capabilities, depends on the difference between the PA saturation power \(P_{\mathrm{sat}}\) and \(P_{\mathrm{avg}}\) as
\begin{equation} \label{avail IBO}
	\mathrm{IBO}_{\mathrm{avail}} = P_{\mathrm{sat}} - P_{\mathrm{avg}}.
\end{equation}
To maintain linear amplification and avoid signal distortion, the system must satisfy
\begin{equation}
	\mathrm{IBO}_{\mathrm{avail}} \geq \mathrm{IBO}_{\mathrm{req}}.
\end{equation}

By introducing the RIS, the effective channel gain is enhanced, enabling a reduction in the required transmit power \(P_{\mathrm{avg}}\) for a given quality-of-service target. This reduction leads to an increase in the available IBO without changing the hardware saturation power \(P_{\mathrm{sat}}\), i.e.,
\begin{equation}
	\mathrm{IBO}_{\mathrm{avail}}^{\mathrm{RIS}} = P_{\mathrm{sat}} - P_{\mathrm{avg}}^{\mathrm{RIS}} > \mathrm{IBO}_{\mathrm{avail}}^{\mathrm{no\,RIS}} = P_{\mathrm{sat}} - P_{\mathrm{avg}}^{\mathrm{no\,RIS}},
\end{equation}
where $P_{\mathrm{avg}}^{\mathrm{RIS}}$ and  $P_{\mathrm{avg}}^{\mathrm{no\,RIS}}$ represent the average transmit power with and without the RIS respectively, with $P_{\mathrm{avg}}^{\mathrm{no\,RIS}} = P_{\mathrm{avg}}^{\mathrm{RIS}} + G_{RIS}$ and $G_{RIS}$ being the equivalent RIS gain.
Therefore, the RIS-assisted system achieves the same linearity requirement in PA with a lower transmission power, reducing power consumption and improving the power amplifier’s operating efficiency. 
This characteristic highlights the advantage of RIS in energy-constrained scenarios, providing a more favorable trade-off between linearity and power efficiency.
\par
In summary, the incorporation of RIS allows for a smaller required IBO in practice or equivalently an increased available IBO for the same transmit power, thereby enhancing system performance and energy efficiency.

\subsection{Complexity Analysis and Scalability}
The introduction of FTN induces colored noise and intentional ISI, creating a coupled effective channel structure.
However, since the pulse shaping is localized, the FTN processing involves banded convolution scaling linearly as $\mathcal{O}(MNW)$, where $W$ is the effective pulse span.
For the RIS configuration, the proposed phase alignment, e.g., Algorithm 1 in Sec. \Rmnum{4}, incurs a low complexity of $\mathcal{O}(PQ)$. 
Crucially, this overhead is amortized over the channel coherence time, as RIS phases are not updated on a per-symbol basis.
Regarding detection, while the adopted LMMSE detector theoretically entails $\mathcal{O}((MN)^3)$ complexity via direct inversion, the effective channel matrix retains a banded-block-Toeplitz structure.
For practical scalability, this structure can be exploited by advanced banded solvers to reduce the complexity to approximately $\mathcal{O}(N_{\mathrm{sys}}W^2)$, where $N_{\mathrm{sys}}\approx MN$ denotes the dimension of the underlying linear system, i.e., the number of DD symbols per OTFS frame.
This ensures that the proposed scheme remains tractable for moderate grid sizes while offering significant spectral efficiency gains.

\begin{algorithm}[t] \label{algorithm}
	\caption{RIS Phase Optimization for RIS-OTFS-FTN with Cascaded Delay-Doppler Channel}
	\label{alg:ris_phase_optimization}
	\begin{algorithmic}[1]
		\State \textbf{Input:} Cascaded DD channel responses $\tilde{\mathbf{H}}_q$ for $q = 1, \dots, Q$
		\State \textbf{Initialize:} Phase shifts $\theta_q \in [0, 2\pi)$ and fixed amplitudes $\beta_q = 1$
		\State Compute the combined unoptimized effective channel by using \eqref{effective channel without phase optimization}
		\State Identify set of non-zero entries $\{ h_{\mathrm{tot}}(k_p, l_p) \}_{p=1}^{P}$ in $\tilde{\mathbf{H}}$
		\For{$p = 1$ to $P$}
		\State Evaluate power metric by using \eqref{strongest path}
		\EndFor
		\State Select strongest path by $p^* = \arg\max_p \text{Power}[p]$
		\State Let $(k^*, l^*) = (k_{p^*}, l_{p^*})$
		\For{$q = 1$ to $Q$}
		\State Compute optimal phase shift by using \eqref{optimal phase shift}
		\State Update phase factor: $\phi_q^{\mathrm{opt}} = e^{j\theta_q^{\mathrm{opt}}}$
		\EndFor
		\State Compute optimized channel by $\mathbf{H}_{\mathrm{opt}} = \sum_{q=1}^{Q} \beta_q \phi_q^{\mathrm{opt}} \cdot \tilde{\mathbf{H}}_q$
		\State \textbf{Output:} Optimized effective channel $\mathbf{H}_{\mathrm{opt}}$
	\end{algorithmic}
\end{algorithm}

\section{RIS Design}

\subsection{Adjusted Phase Algorithm}
In our work, the goal of RIS phase optimization is to adjust the phase shifts of each RIS reflecting element to maximize the effective channel gain at the receiver.
A key characteristic of an RIS system is its capability to dynamically reconfigure reflection coefficients in accordance with channel conditions. 
In OTFS modulation, the channel is represented in the DD domain, where variations occur at a slower rate compared to the time-frequency domain representation. 
Consequently, the phase adjustment frequency in the RIS can be reduced.
Here, we consider the discrete phase adjustment, of which the optimization procedure is described below.
\par
According to \eqref{vectorized received signals}, we have the cascaded DD channel response of the $q$-th RIS element be denoted by $\tilde{\mathbf{H}}_q$.
Thus, the combined effective channel with phase optimization is expressed as
\begin{align} \label{effective channel without phase optimization}
	\tilde{\mathbf{H}}_q=\sum_{q=1}^Q{\beta _qe^{j\theta _{q}^{v}}\tilde{\mathbf{H}}_q},
\end{align}
where amplitude attenuation $\beta _q =1$ is defined in this section for better analysis,
$\{\theta_q^v\}_{q=1,v=1}^{Q,V}$ denote a set of discrete adjusted phases with $q \in \{1, 2, \dots, Q\}$ and $v \in \{1, 2, \dots, V\}$, each $\theta_q^v \in [0, 2\pi)$, and the adjusted-phase set is $\{\theta_q^1, \theta_q^2, \dots, \theta_q^V\}$ for each $q$,
\par
Let $\{h_{tot}(k_p,l_p)\}_{p=1}^{P}$ denote the non-zero entries in $\tilde{\mathbf{H}}_q$ corresponding to $P$ cascaded propagation paths. Among them, the strongest path is selected by maximizing the following metric given as follows:
\begin{align} \label{strongest path}
	p^* = \arg \max_{p \in \{1,\dots,P\}} \left| h_{\mathrm{tot}}(k_p, l_p) + \sum_{i=1}^{L} h^{(i)}(k_p, l_p) \right|^2.
\end{align}
To align the phase of each RIS element constructively on the selected tap $(k^*, l^*)$ of the strongest path, the optimal phase shift is computed as
\begin{align} \label{optimal phase shift}
	\theta_i^{\mathrm{opt}} = \arg\left( h_{\mathrm{tot}}(k^*, l^*) \right) - \arg\left( h^{(q)}(k^*, l^*) \right),
\end{align}
with $\arg(\cdot)$ representing the argument of a complex number, and the corresponding optimized phase factor given by $\phi_i^{\mathrm{opt}} = e^{j \theta_i^{\mathrm{opt}}}$.
Therefore, after resulting optimized effective channel as $\mathbf{H}_{\mathrm{opt}} = \sum_{i=1}^{L} \phi_i^{\mathrm{opt}} \cdot \mathbf{h}^{(i)}$, we can respectively obtain the received signals in time and DD domain as
\begin{align} \label{optimal phase in time and DD domain}
	&y_{\mathrm{opt}}(t)=\sum_{q=1}^Q{\sum_{p_1=0}^{P_1-1}{\sum_{p_2=0}^{P_2-1}{h_{q,p_{1,2}}^{\mathrm{eff}}}\beta _qe^{j\left( \theta_1 -\theta _{i}^{\mathrm{opt}} \right)}s\left( t-\tau _{q,p_{1,2}}^{\mathrm{eff}} \right)}},  \nonumber
	\\
	&y[a]=\sum_{q=1}^Q{\sum_{p_1=0}^{P_1-1}{\sum_{p_2=0}^{P_2-1}{h_{q,p_{1,2}}^{\mathrm{eff}}\beta _q\,e^{j\left( \theta _2-\theta _{i}^{\mathrm{opt}} \right)}s\left[ [a-\varepsilon _{q,p_{1,2}}]_{NM} \right]}}}
\end{align}
with $\theta_1 =2\pi \nu _{q,p_{1,2}}^{\mathrm{eff}}(t-\tau _{q,p_{1,2}}^{\mathrm{eff}})$ and $\theta _2=2\pi \frac{k_{q,p_{1,2}}(a-\varepsilon _{q,p_{1,2}})}{NM}$.
Detail steps for optimization can be found in Algorithm 1.
\remark{According to \eqref{spectral efficiency}, the spectral efficiency is expected to improve by employing the RIS phase optimization strategy. 
This is because optimizing the RIS phase shifts enhances the effective channel gain by aligning the reflected signals constructively at the receiver. 
Thus, the matrix product $\mathbf{H}_{\mathrm{eff}}\mathbf{H}_{\mathrm{eff}}^{H} $ contains stronger eigenvalues, leading to a larger determinant inside the logarithm. 
Although random phase adjustment may also increase compared to the case without RIS, which is because random phase configurations can still introduce diversity and enable some constructive interference at the receiver by chance, the effective channel gain $\mathbf{H}_{\mathrm{eff}}$ is generally weaker than in the case of optimal phase alignment \textcolor{green}{since the phase shifts are not aligned to maximize the received signal power.}
Consequently, the performance gain from random phase modulation is typically limited and less consistent.
More corresponding results and insights are given in Sec. \Rmnum{5}.
}
\remark{Identifying the strongest path from $P=P_1+P_2$ dominant multipath components require $\mathcal{O}(P)$ operations. 
	Subsequently, calculating the optimal phase shifts for $Q$ reflection elements involve $\mathcal{O}(Q)$ complex additions and phase extractions. 
	Therefore, the total computational complexity is in the order of $\mathcal{O}(P + Q)$. 
	Compared to the signal detection complexity, e.g., $\mathcal{O}( (NM)^3 )$ for LMMSE, the overhead introduced by the RIS phase design can be negligible.}

\subsection{Robustness to Channel Estimation Errors}
\label{subsec:ICSI}
In practical high-mobility scenarios with RIS applications, obtaining perfect channel state information (CSI) is challenging. 
To evaluate the robustness of the proposed RIS phase alignment strategy, we model the impact of imperfect CSI. 
For better presentation, we assume the transmitter-to-RIS link, being stationary, is accurately estimated. 
However, the RIS-to-User link suffers from estimation errors due to user mobility and difficulty of estimating reflecting link. 
The estimated channel impulse response with imperfect CSI for the RIS-to-User link is modeled as
\begin{equation}
	\hat{\boldsymbol{h}}_{q}^{2} = \boldsymbol{h}_{q}^{2} + \mathbf{e}_q,
\end{equation}
where $\boldsymbol{h}_{q}^{2}$ denotes the vector of path gains $[h_{q,0}^{2}, \dots, h_{q,P_2-1}^{2}]^T$ corresponding to the time-domain response in \eqref{h1 T2}, and $\mathbf{e}_q \sim \mathcal{CN}(\mathbf{0}, \delta_{e^2}^2 \mathbf{I})$ represents the channel estimation error vector. 

\subsection{Impact of Finite Resolution Phase Quantization}
\label{subsec:Quantization}
Practical RIS implementations are constrained by hardware limitations, typically supporting only a finite set of discrete phase levels. 
To evaluate the impact of this practical constraint, we consider a $B$-bit quantization scheme. 
The set of feasible discrete phase shifts is defined as $\mathcal{F}_B = \{0, \frac{2\pi}{2^B}, \dots, \frac{2\pi(2^B-1)}{2^B}\}$. 
Consequently, the continuous optimal phase $\theta_q^{\mathrm{opt}}$ obtained from Algorithm 1 is quantized to the nearest discrete level $\bar{\theta}_q \in \mathcal{F}_B$ .
The resulting quantized phase shift matrix, denoted as $\bar{\mathbf{\Phi}} = \mathrm{diag}(\beta_1 e^{j\bar{\theta}_1}, \dots, \beta_Q e^{j\bar{\theta}_Q}) \otimes \mathbf{I}_{NM}$, is then applied to the effective channel construction. This quantization introduces a phase mismatch that degrades the passive beamforming gain. The trade-off between the quantization resolution $B$, e.g., 1, 2, and 3-bit, and the system error performance is numerically investigated in Section V.

\begin{figure*}
	\centering
	\includegraphics[width=16.5cm,height=9cm]{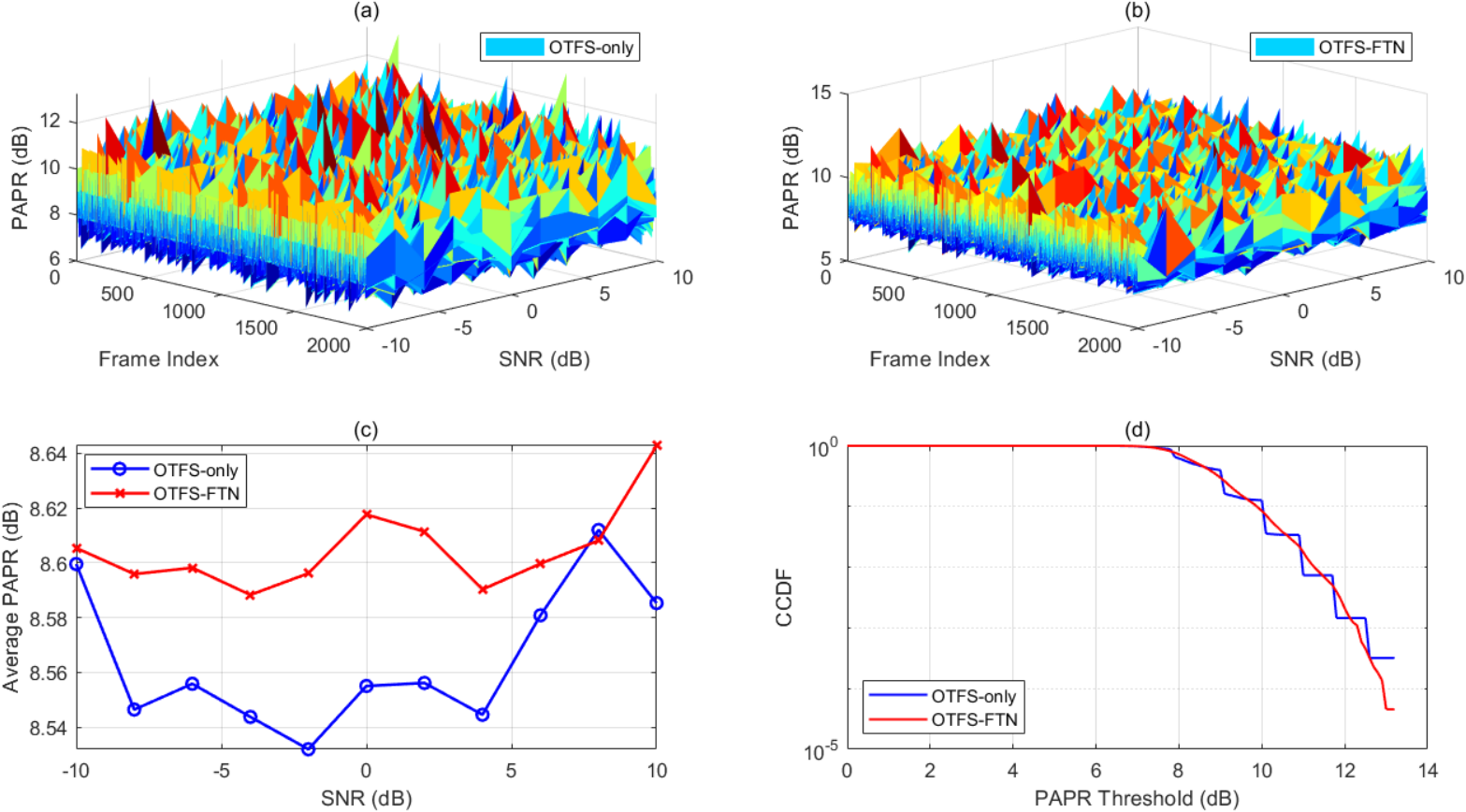}\\
	\caption{PAPR (a \& b) and CCDF (c \& d) of the OTFS-FTN and OTFS schemes with $\alpha=0.8$, $M=32$ and $N=32$.}
	\label{fig:PAPR_CCDF}
\end{figure*}

\begin{figure}
	\centering
	\includegraphics[width=8cm,height=7cm]{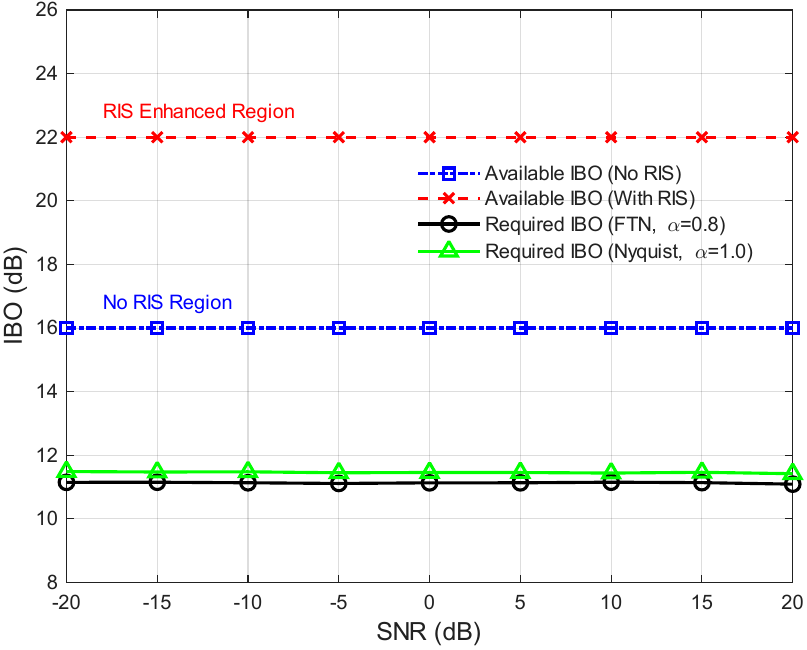}\\
	\caption{IBO of the RIS-OTFS-FTN and OTFS-FTN schemes with $\alpha=0.8$ and $1$, $M=32$ and $N=32$.}
	\label{fig:IBO}
\end{figure}

\begin{figure}
	\centering
	\includegraphics[width=8cm,height=8cm]{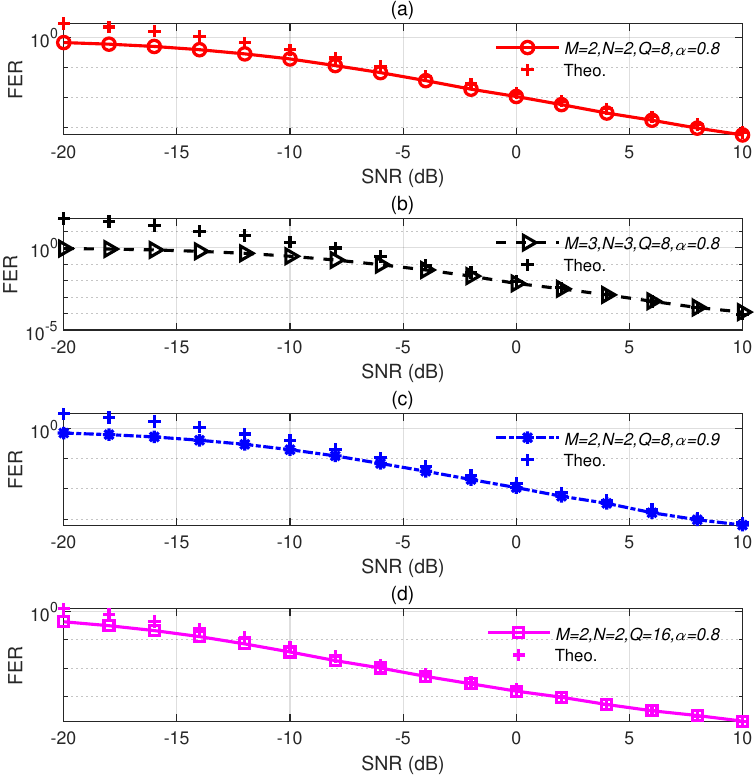}\\
	\caption{Small-scale theoretical and simulation FER of the RIS-OTFS-FTN scheme with different $M$, $N$, $Q$, and $\alpha$.}
	\label{fig:FER}
\end{figure}

\begin{figure}
	\centering
	\includegraphics[width=8cm,height=11cm]{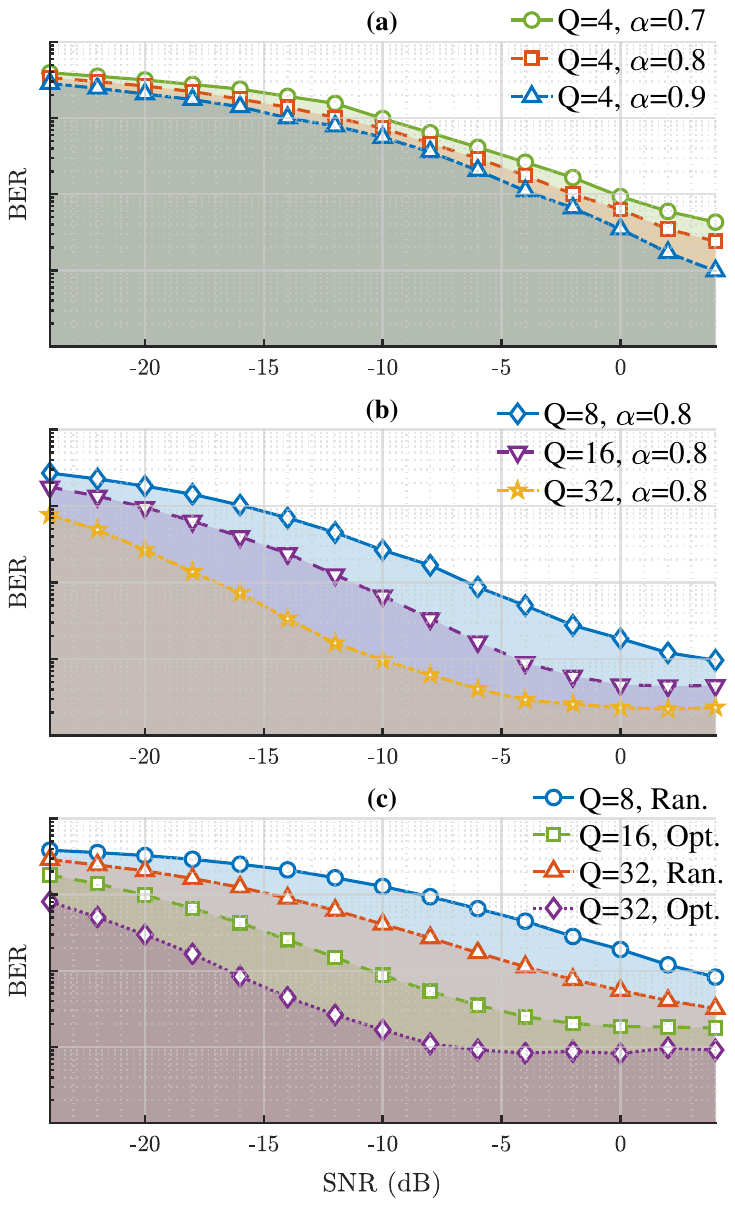}\\
	\caption{BER of the MMSE detector-assisted RIS-OTFS-FTN scheme with various comparing cases: (a) $Q=4$ and $\alpha=0.7, 0.8, 0.9$; (b) $\alpha=0.8$ and $Q=8, 16, 32$; (c) $\alpha=0.8$,$Q=8, 16, 32$ in optimal phase adjustment (Opt.) and the random phase one (Ran.).}
	\label{fig:BER_MMSE}
\end{figure}

\begin{table}[!]
	\centering
	\small
	\caption{Delay--Power Profile of the EVA Channel Model}
	\begin{tabular}{c c c c}
		\hline
		\textbf{Path Index} & \textbf{Delay (ns)} & \textbf{Power (dB)} & \textbf{Linear Power} \\
		\hline
		1 & 0    & 0.0    & 0.1964 \\
		2 & 30   & -1.5   & 0.1389 \\
		3 & 150  & -1.4   & 0.1433 \\
		4 & 310  & -3.6   & 0.0870 \\
		5 & 370  & -0.6   & 0.1709 \\
		6 & 710  & -9.1   & 0.0246 \\
		7 & 1090 & -7.0   & 0.0391 \\
		8 & 1730 & -12.0  & 0.0124 \\
		9 & 2510 & -16.9  & 0.0045 \\
		\hline
	\end{tabular}
	\label{tab:EVA_profile}
\end{table}

\vspace{-10pt}
\section{Simulation Results}
In this section, theoretical and simulation results of the RIS-OTFS-FTN scheme, e.g., PAPR, CCDF, IBO, FER, BER, and spectral efficiency are provided.
Here, to approach the channel environment in suburban or urban high-speed scenarios, a standard EVA wireless channel model is employed in this section \footnote{EVA channel model is a representative wireless fading model widely employed for simulating high-mobility environments, such as those encountered in LTE, 5G, and performance evaluations of advanced schemes like the RIS-OTFS-FTN system. As an extended version of the Vehicular A (VA) model recommended by the 3GPP and ITU standards \cite{ref35}, the EVA model features a longer delay spread and more complex multi-path characteristics. These enhancements allow it to more accurately capture the impact of multi-path interference in high-speed scenarios.}.
The detail values of simulation parameter in EVA model are given in Table \ref{tab:EVA_profile}.
For a clear presentation and comparison in this section, several other parameters are also fixed, including size of QAM $M_{QAM}=2$, carrier frequency $f_c=4$ GHz, subcarrier spacing $\bigtriangleup f=15$ KHz, unit transmitted energy of each symbol $E_x=1$, maximum speed for user $v_{max}=120$ km/h, and SNR as $\frac{_{E_x}}{^{\sigma ^2}}$.
Two parameters of RRC filter are firstly given, e.g., RRC roll-off factor $\beta_{RRC}=0.25$ and symbol period span $\mathrm{s}_{RRC}=8$.
Moreover, Monte Carlo simulations are conducted with each experiment performed over $3\times 10^3$ independent channel realizations, while considering the large complexity in calculation and running simulation.
In consideration of the practical implementation of an RIS, discrete phase shifts with 3 bits allocated to each are adopted for simulation purposes, and the set of feasible values is expressed as $\theta _{q}^{v}\in \left\{ 0,\frac{2\pi}{8},\frac{4\pi}{8}, \frac{6\pi}{8},\pi ,\frac{10\pi}{8},\frac{12\pi}{8},\frac{14\pi}{8} \right\} $.
The results are presented under the assumption that there exists no direct vehicle-user channel, with the RIS positioned equidistantly from both the vehicle and the user.
\par  
\figref{fig:PAPR_CCDF} compares the PAPR characteristics of the proposed RIS-OTFS-FTN scheme with those of the conventional OTFS scheme. 
Note that the RIS is not considered in this figure since the PAPR is only evaluated the transmitted signals, so $Q$ can be ignored here and the legend of our scheme in \figref{fig:PAPR_CCDF} is simplified as the OTFS-FTN.
The results reveal that the FTN-based scheme exhibits a PAPR distribution closely matching that of the Nyquist-based OTFS counterpart, indicating that for the current parameter settings, the application of FTN exerts an acceptable influence on the overall PAPR behavior. 
(c) and (d) of \figref{fig:PAPR_CCDF} present the CCDF of the PAPR for both schemes. 
The two CCDF curves almost completely overlap, demonstrating that the FTN-enabled scheme yields negligible PAPR penalty compared to conventional OTFS. 
In the PAPR range of approximately $8$--$12$~dB, the red curve (FTN) is slightly shifted to the right and lies marginally above the blue curve (OTFS), with minor fluctuations in some segments. 
This behavior indicates that FTN marginally increases the probability of higher PAPR occurrences in the tail region, although the increase is minimal. 
From a practical perspective, the observed PAPR behavior implies that the spectral efficiency advantages of FTN can be leveraged without incurring significant PAPR-related drawbacks. Moreover, as later illustrated in \figref{fig:PAPR_CCDF}, the deployment of RIS compensates for even this small PAPR rise by enhancing the effective channel gain, thereby enabling reduced transmit power while meeting the same quality-of-service constraints. Therefore, despite the minimal PAPR difference, the proposed RIS-OTFS-FTN scheme maintains clear advantages in spectral efficiency and energy efficiency in high-mobility EVA channel scenarios.
\par 
\figref{fig:IBO} illustrates the IBO performance of the proposed RIS-OTFS-FTN scheme compared to that of the OTFS-FTN scheme without RIS assistance. 
Notably, considering normal configuration in the RIS and better presentation, a fixed value of $G_{RIS}=6$ dB is given here.
Here, the available IBO is computed according to \eqref{avail IBO}, where the PA saturation power is fixed and the average transmit power is determined based on $E_f$. 
As shown in the figure, the RIS-assisted scheme consistently achieves a higher available IBO across the entire SNR range, with the improvement becoming more pronounced at higher SNR values. 
This behavior is attributed to the additional channel gain introduced by the RIS, which enables the transmitter to operate at a lower average power for the same error performance, thereby increasing the linear margin  \footnote{linear margin refers to the margin of input signal dynamic range that a system or device can withstand while maintaining the linear characteristics of the output signal. To put it simply, it is the allowable fluctuation range of input power without causing significant nonlinear distortions (such as harmonics and inter-modulation interference). A larger margin indicates a stronger ability of the system to resist distortion when dealing with signal changes.} between the average transmit power and the PA saturation limit.
The results clearly demonstrate the energy-efficiency benefits of incorporating RIS in high-mobility OTFS-FTN systems. 
In particular, the increased available IBO relaxes the linearity requirements of the PA, reduces the risk of nonlinear distortion, and allows for more efficient operation without compromising signal integrity. 
From a system design perspective, this implies that the RIS-OTFS-FTN scheme can maintain robust performance even when operating closer to the PA’s nonlinear region, thus reducing hardware constraints and improving overall power efficiency. 
Moreover, the observed IBO advantage complements the PAPR analysis in \figref{fig:PAPR_CCDF}, as the RIS gain not only offsets the minimal PAPR increase due to FTN but also extends the operational flexibility of the transmitter in practical deployments.
Moreover, by comparing the results with Nyquist signaling baseline ($\alpha=1$, i.e., without FTN) alongside the proposed FTN scheme ($\alpha=0.8$), it is observed that the curve for FTN ($\alpha=0.8$) is comparable to, and in some regions even marginally lower than, that of the Nyquist baseline. 
That is, the proposed scheme achieves higher spectral efficiency without incurring a penalty in terms of PAPR or linearity requirements. 
\par  
For \figref{fig:FER}, the theoretical and simulation FER results are presented with a small-scale running in different $M$, $N$, and $\alpha$, as given in figure.
\footnote{The use of small-scale grids (e.g., $M, N \leq 3$) in these evaluations is necessitated, due to the extremely large computational complexity of constructing and evaluating all possible symbol combinations and ergodic calculation in the ML detector. }
The simulation results are still under the assumption of cascaded channel gain $\mathbf{H}_{\mathrm{eff}}$ following the Rayleigh fading, and the RIS gain is also considered.
In \figref{fig:FER}, we can observe that the theoretical upper bound on FER closely aligns with the simulation results, particularly at higher SNR values where SNR $\gg -5$. 
A slight divergence is observed at lower SNR, attributable to the looseness of the union bound in regimes dominated by noise. 
Comparing to (a), (c), and (d), (b) increases the grid size to $M=3$, $N=3$ while maintaining $Q=8$ and $\alpha=0.8$, revealing a marginal degradation in FER performance. 
This is expected, as larger grids introduce more symbols per frame, increasing susceptibility to ISI and channel impairments, though the theoretical curve continues to serve as a reliable upper bound. 
(c) examines the impact of a higher compression factor $\alpha=0.9$ with $M=2$, $N=2$, and $Q=8$, demonstrating improved FER relative to $\alpha=0.8$ in (a). The reduced FTN compression mitigates ISI, leading to lower error rates, with the theoretical and simulated curves exhibiting strong agreement. 
Finally, (d) doubles the RIS elements to $Q=16$ while keeping $M=2$, $N=2$, and $\alpha=0.8$, yielding a notable enhancement in FER performance over (a). 
The additional RIS elements provide greater received SNR, effectively strengthening the cascaded channel and reducing error probabilities, as corroborated by the tighter alignment between theory and simulation.
\par  
\figref{fig:BER_MMSE} illustrates the BER performance of the MMSE detector-assisted RIS-OTFS-FTN scheme across same values of $M=16$ and $N=16$ and various configurations in $Q$ and $\alpha$, as given in title.
Note that the Opt. and Ran. in legends represent the optimal RIS phase design in this paper and random phase selection, respectively.
In (a), with a fixed number of RIS elements $L=4$ and different $\alpha=0.7, 0.8, 0.9$, the BER decreases as $\alpha$ increases, approaching the Nyquist signaling regime ($\alpha=1$). 
This is also because higher $\alpha$ values reduce the severity of intentional ISI introduced by FTN, leading to improved detection accuracy. 
However, the differences in BER among these $\alpha$ values are relatively approached, particularly at low-to-moderate SNR, indicating that the spectral efficiency gains from lower $\alpha$ come at a limited cost in error performance under the MMSE detector.
(b) examines the impact of increasing reflection elements in the RIS ($L=8, 16, 32$) with fixed $\alpha=0.8$. 
As $Q$ increasing, the BER performance improves significantly due to enhanced channel gain through the RIS. 
Notably, an error floor trend emerges at higher SNR, becoming more pronounced and appearing earlier at lower SNR with larger $Q$. 
This behavior is attributed to the increased ISI in the cascaded vehicle-RIS-user channels, where more reflection elements amplify multi-path components, exacerbating interference in the DD domain and limiting the MMSE detector's ability to fully mitigate it.
For (c), the Ran. design yields substantially higher BER across all SNR by comparing to the Opt. design with differences exceeding an order of magnitude in some regions, especially with high $Q$ and in low SNR range.
This underscores the critical role of  Algorithm 1, which aligns reflected signals constructively to maximize channel gain.
\par
\begin{figure}
	\centering
	\includegraphics[width=8cm,height=6cm]{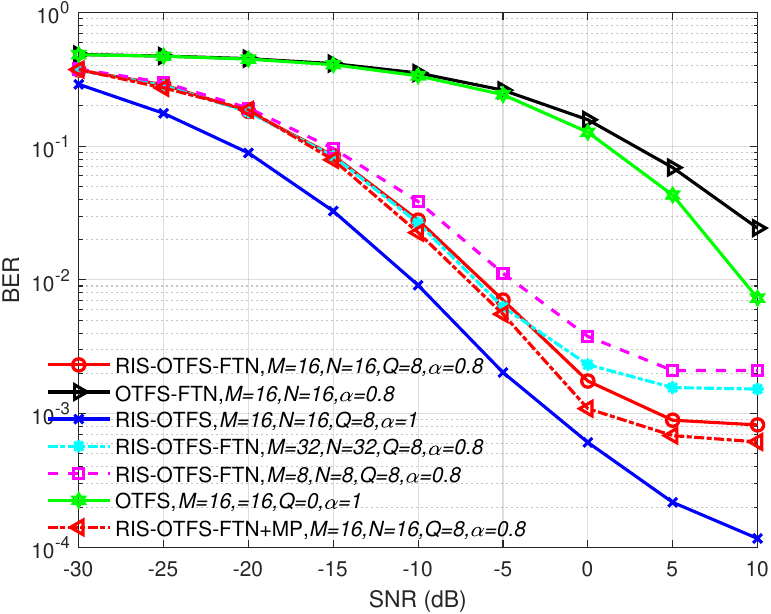}\\
	\caption{BER of the RIS-OTFS-FTN scheme with/without RIS, FTN, MP detector, as well different $M$ and $N$.}
	\label{fig:BER_RIS}
\end{figure}

  
\figref{fig:BER_RIS} depicts the BER performance of the RIS-OTFS-FTN scheme with MMSE detector in comparison to benchmark variants, including schemes without RIS or FTN, as well as different $\alpha$ $M$ and $N$., as shown in figure.
Notably, the RIS-OTFS scheme ($\alpha=1$) achieves superior performance relative to FTN-enabled variants at equivalent grid sizes, attributable to its use of ideal rectangular pulses rather than the RRC filter employed in FTN processing. 
And the RIS-OTFS-FTN with $M=16$, $N=16$, $Q=8$, and $\alpha=0.8$ extremely outperforms the OTFS-FTN without RIS, demonstrating that RIS can better enhance the effective channel strength and mitigates fading in high-mobility scenarios. 
Similarly, comparing RIS-OTFS-FTN to RIS-OTFS, FTN variant introduces additional ISI due to time-domain compression, leading to a performance degradation, particularly evident at higher SNR where an error floor emerges, which is because increased grid dimensions amplify the impact of residual ISI and channel dispersion in the DD domain, limiting LMMSE detector's equalization capability.
Besides, the BER performance improves as $M$ and $N$ increases, which happens because larger OTFS grids offer finer resolution in the DD domain, enabling better exploitation of channel diversity and more effective equalization of multi-path and Doppler effects in detection.
For the results with message passing (MP) detector, which has 100 maximum number of MP detector iterations and damping parameter $\omega_{MP}=0.8$, we observe that despite the dense channel graph, the MP detector's ability to explicitly model the FTN interference structure allows it to outperform LMMSE in the high-SNR regime, effectively improving the error performance in high SNR range. This confirms that while MP is computationally expensive in RIS environments, it is necessary for maximizing reliability when aggressive FTN compression is applied. 
\par
\begin{figure}
	\centering
	\includegraphics[width=8cm,height=11cm]{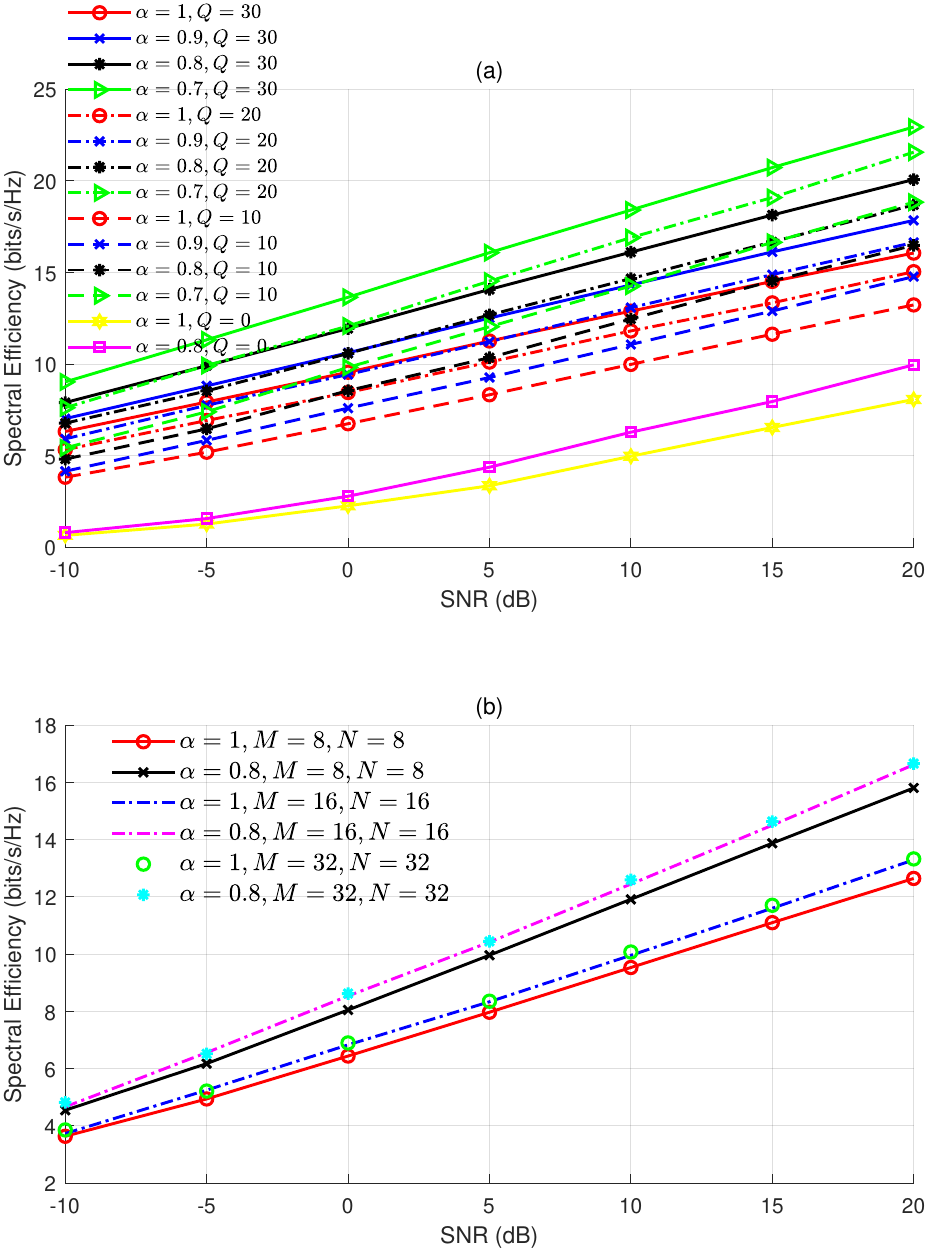}\\
	\caption{Spectral efficiency of the RIS-OTFS-FTN scheme with different cases in (a) $\alpha=1, 0.9, 0.8, 0.7$, $Q=30, 20, 10, 0$, $M=16$, and $N=16$; (b) $\alpha=1, 0.9, 0.8, 0.7$, $Q=10$, $M=8, 16, 32$, and $N=8, 16, 32$.}
	\label{fig:SE}
\end{figure}

  
\figref{fig:SE} evaluates the spectral efficiency of the RIS-OTFS-FTN under various configurations, including same $Q=32$, different $\alpha$,  $M$, and $N$, as given in title and legend of figure. 
In (a), with fixed $M=16$ and $N=16$, the spectral efficiency is compared for $\alpha = 1, 0.9, 0.8, 0.7$ across $Q = 30, 20, 10, 0$. 
As $\alpha$ decreases, the spectral efficiency exhibits a significant increase, consistent with the inverse proportionality to $\alpha$ in the normalization factor of \eqref{spectral efficiency}. 
Notably, as shown in \figref{fig:BER_RIS}, while lower $\alpha$ values degrade BER due to heightened ISI, the spectral efficiency gains are substantial and continue to grow with diminishing $\alpha$, highlighting a clear trade-off between error performance and throughput. 
Furthermore, increasing $Q$ amplifies the spectral efficiency across all $\alpha$, attributable to the RIS-induced boost in effective channel gain, particularly at higher SNR where the scheme approaches capacity.
\figref{fig:SE} (b) fixes $Q=10$ and varies $\alpha = 1, 0.9, 0.8, 0.7$ with $M=N=8, 16, 32$. 
Similar to (a), spectral efficiency increases with decreasing $\alpha$. 
Intriguingly, smaller grid sizes in $M N$ yield higher spectral efficiency, as the normalization in \eqref{spectral efficiency} scales inversely with $M N$, implying that compact frames benefit more from FTN compression relative to the occupied time-frequency resources. 
Overall, the RIS-OTFS-FTN scheme embodies a strategic trade-off in high-mobility wireless environments: FTN enhances SE at the cost of BER degradation, RIS scaling bolsters both reliability and efficiency, and grid sizing influences the normalized performance. 
By considering these effects, the proposed framework offers a versatile alternative to conventional OTFS, adaptable to diverse scenarios such as spectrum-constrained urban vehicular networks or energy-limited high-speed links, thereby providing an additional tool to combat challenging propagation conditions.
\textcolor{red}{In \figref{fig:SE} (b), impact of the grid size on spectral efficiency under a constant average transmit power, revealing a fundamental trade-off governed by two competing mechanisms. 
For excessively compact grids, e.g., $M=N=8$, the discrete OTFS block suffers from severe boundary truncation, where uncaptured multipath energy spills over symbol boundaries, leading to a noticeable SE degradation. 
Conversely, when the grid is enlarged, e.g., from $16 \times 16$ to $32 \times 32$, to encapsulate almost all multipath energy without truncation, the SE growth saturates with negligible differences. 
This upper-bound saturation is rigorously justified by \eqref{spectral efficiency}. 
While enlarging the grid beyond the channel's maximum delay spread conserves total received energy, it simultaneously allows the FTN-induced colored noise to diffuse across a broader dimensional subspace.  
Detailed proof of this conclusion is in Appendix B.}
\par
\textcolor{red}{A joint analysis is imperative to identify the practical system operating region. Table \ref{tab:joint_se_ber} extracts this joint performance at SNR = 10 dB under the recommended $16 \times 16$ grid. By employing FTN ($\alpha=0.8$) with $R_s$ as the baseline Nyquist symbol rate ($1/T_0$), the system accelerates the throughput to $1.25R_s$, leading to an inherent increase in SE. However, in such a high-rate non-orthogonal transmission mode, the signal detection becomes more sensitive to the operating SNR environment. Thus, without RIS assistance, the baseline OTFS-FTN scheme exhibits a BER of $2.44 \times 10^{-2}$ at the given SNR. Integrating a moderate RIS array ($Q \approx 10$) optimizes this performance boundary. The massive passive array gain provides substantial SNR headroom, effectively compensating for the reliability pressure induced by the accelerated symbol rate. Thus, the RIS-OTFS-FTN scheme successfully sustains the $1.25R_s$ rate, achieving a remarkably high SE of 12.46 bits/s/Hz while maintaining a robust BER of $8.24 \times 10^{-4}$. This confirms that the integration of RIS is a strategic enabler for deploying high-efficiency FTN signaling in practical scenarios.}
\par
\begin{figure}
	\centering
	\includegraphics[width=8cm,height=6cm]{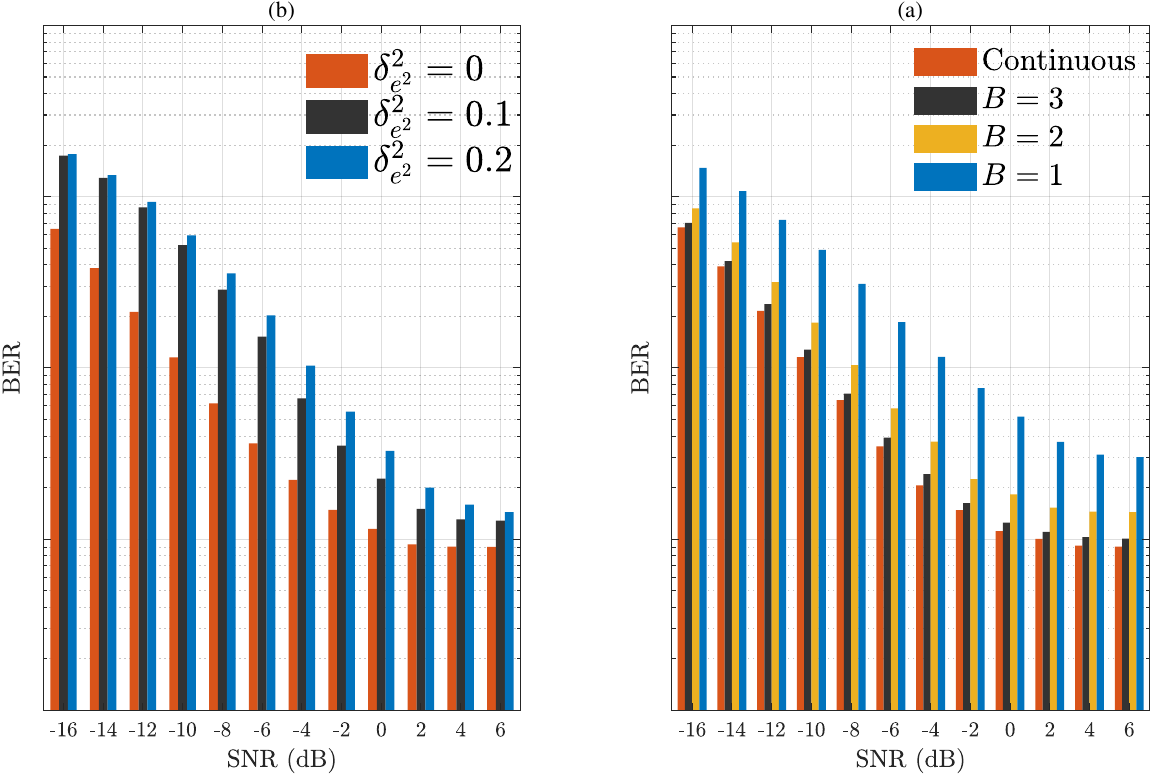}\\
	\caption{BER of the RIS-OTFS-FTN scheme with imperfect CSI in different $\delta_{e^2}^2$ and varying number of quantized adjusted phases $B$.}
	\label{fig:BER_ICSI_quan}
\end{figure}

\figref{fig:BER_ICSI_quan} depicts the BER performance under imperfect CSI with varying estimation error variances $\sigma_e^2 \in \{0, 0.1, 0.2\}$, and investigates the impact of finite phase resolution on the BER performance.
It is observed that while the performance naturally degrades as the estimation accuracy decreases, the system does not suffer from catastrophic failure.
Even with a relatively large error variance of $\sigma_e^2=0.2$, the proposed RIS phase alignment strategy maintains a valid gain, exhibiting a graceful performance degradation. 
This confirms the robustness of the proposed scheme against system uncertainties in high-mobility scenarios.
(b) evaluates the impact of phase quantization resolution $B \in \{1, 2, 3\}$ bits compared to the ideal continuous phase case.
The results demonstrate that the performance gap between the $B=3$ bits quantization and the continuous phase is negligible.
Also, the $B=2$ bits configuration incurs only a marginal SNR penalty, approximately 1-2 dB at $\text{BER}=10^{-2}$, while significantly reducing hardware complexity.
However, the 1-bit quantization results in a noticeable performance loss due to the large phase mismatch.
These results indicate that a low-resolution RIS architecture is sufficient to realize the benefits of the proposed RIS-OTFS-FTN scheme, making it a cost-effective solution for practical deployment. 
\par
\begin{figure}
	\centering
	\includegraphics[width=9cm,height=5cm]{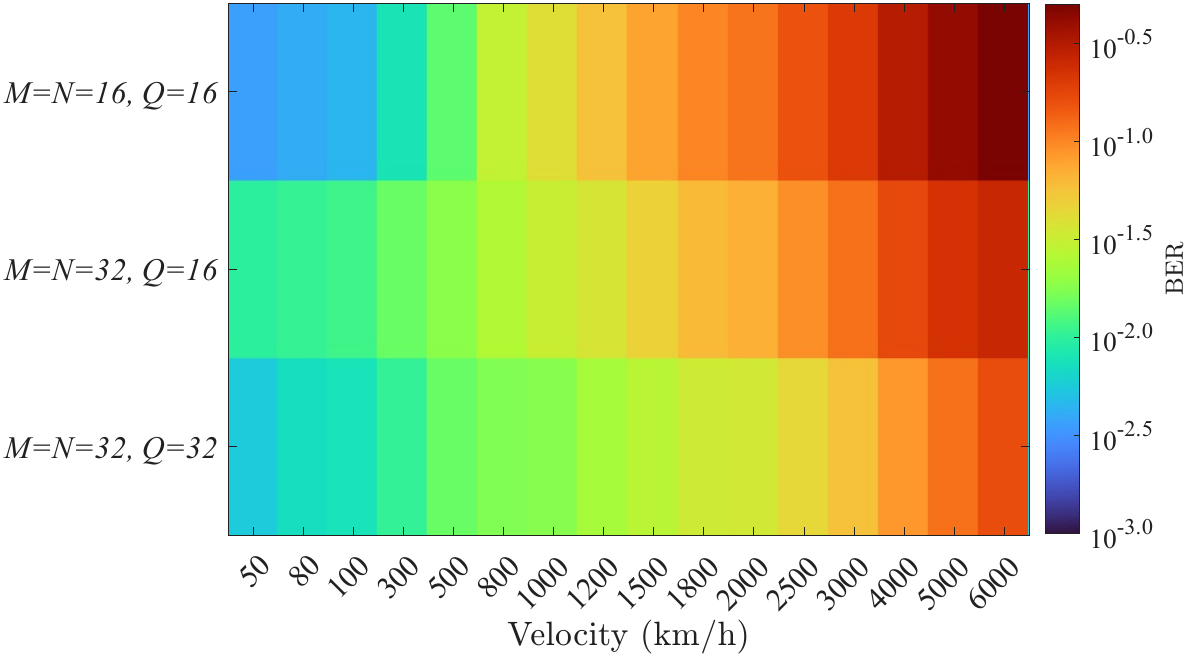}\\
	\caption{Operating limits analysis of the RIS-FTN-OTFS: BER versus User Velocity (km/h) at SNR = 20 dB.}
	\label{fig:velocity_limit}
\end{figure}

\figref{fig:velocity_limit} illustrates the BER performance as a function of user velocity ranging from $0$ to $6000$ km/h, with a fixed high SNR of $20$ dB to isolate the impact of Doppler-induced impairments.
The "operating limit" is defined as the maximum velocity.
For the configuration with $M=N=16$ and $Q=16$, the system exhibits robust performance up to approximately $1,700$ km/h. 
This indicates that for standard high-speed vehicular or railway scenarios ($<500$ km/h), the proposed scheme operates well within its safety margin with negligible Doppler-induced degradation.
Increasing the OTFS grid size to $M=N=32$ significantly extends the tolerance limit to roughly $2,800$ km/h. 
The finer resolution in the Delay-Doppler domain allows for more accurate channel resolvability and effective separation of high-Doppler paths.
Further increasing the number of RIS elements to $Q=32$ pushes the operating limit beyond $4,800$ km/h. The enhanced passive beamforming gain provided by the larger RIS array effectively compensates for the SNR degradation caused by severe inter-carrier interference at hypersonic speeds.
Increasing the OTFS grid size to $M=N=32$ significantly extends the tolerance limit to roughly $2,800$ km/h. 
Besides, A larger $N$ reduces the Doppler bin spacing ($\Delta\nu = \frac{1}{NT}$), enabling the receiver to resolve closely spaced high-Doppler paths that would otherwise merge or cause severe inter-carrier interference (ICI) in a coarser grid, which is why the BER gets better with increase of $MN$.
In summary, the proposed architecture demonstrates exceptional robustness, making it a viable candidate not only for terrestrial high-mobility networks but also for future non-terrestrial networks (NTN) such as LEO satellite communications.  

\begin{table}[!]
	\centering
	\small 
	\setlength{\tabcolsep}{1.8pt} 
	\caption{\textcolor{red}{Joint SE-BER Performance at SNR = 10 dB with $16 \times 16$ grid and LMMSE detector.}}
	\label{tab:joint_se_ber}
	\begin{tabular}{@{}l c c c c@{}}
		\toprule
		\textbf{Scheme} & \textbf{Parameters} & \makecell{\textbf{Symbol} \\ \textbf{Rate}} & \textbf{BER} & \makecell{\textbf{SE} \\ \textbf{(bits/s/Hz)}} \\
		\midrule
		OTFS (Baseline) & $\alpha=1.0, Q=0$ & $R_s$ & $7.30 \times 10^{-3}$ & $4.96$ \\
		OTFS-FTN & $\alpha=0.8, Q=0$ & $1.25 R_s$ & $2.44 \times 10^{-2}$ & $6.28$ \\
		RIS-OTFS & $\alpha=1.0, Q \approx 10$ & $R_s$ & $1.17 \times 10^{-4}$ & $9.98$ \\
		RIS-OTFS-FTN & $\alpha=0.8, Q \approx 10$ & $1.25 R_s$ & $8.24 \times 10^{-4}$ & $12.46$ \\
		\bottomrule
	\end{tabular}
\end{table}

\vspace{-10pt}
\section{Conclusion}
This study has established a comprehensive theoretical framework for the RIS-assisted OTFS-FTN paradigm in next-generation high-mobility wireless ecosystems.
Closed-form expressions for AFER, spectral efficiency, PAPR, and IBO have been rigorously derived and extensively validated through abundant Monte-Carlo simulations, offering deeper insights into the interplay among FTN compression, OTFS modulation, and RIS applications.
Furthermore, a capacity-maximizing RIS phase-optimization algorithm that constructively aligns the cascaded DD channel has been devised at the receiver.
The synergistic exploitation of RIS, OTFS, and FTN has demonstrated substantial enhancements in spectral and energy efficiency while preserving robust error performance under realistic EVA channels.
It should be emphasized that, when it comes to practical deployment, the trade-off among $\alpha$, grid size $M\times N$, and $L$ needs to be carefully balanced.  
As an optional scheme, the RIS-OTFS-FTN framework is proven to have great potential for future vehicular, railway, and satellite networks that demand high throughput under severe Doppler and blockage conditions.

\begin{appendices}
	\section{}
\textcolor{red}{
Following the continuous-time energy formulation for FTN signaling \cite{ref27}, the expected total energy of an OTFS-FTN frame in the continuous-time domain is given by:
\begin{align} \label{frame_energy}
	E_f &= \mathbb{E} \left[ \int_{-\infty}^{\infty} |s(t)|^2 dt \right] \nonumber \\
	&= \mathbb{E} \left[ \sum_{k=0}^{MN-1} \sum_{m=0}^{MN-1} s_k s_m^* g((k-m)T_f) \right] \nonumber \\
	&= \sum_{k=0}^{MN-1} \sum_{m=0}^{MN-1} \mathbb{E} [s_k s_m^*] g((k-m)T_f),
\end{align}
where $g(t) = g_{tx}(t) * g_{rx}^*(-t)$ is the equivalent pulse shaping filter. Considering that the transmitted data symbols are independent and identically distributed (i.i.d.) with zero mean and average energy $\mathbb{E}[|s_k|^2] = E_x$, their mutual correlation satisfies the Kronecker delta property:
\begin{align} \label{1111}
	\mathbb{E}[s_k s_m^*] = 
	\begin{cases} 
		E_x, & k=m \\ 
		0, & k \neq m 
	\end{cases}.
\end{align}
Substituting \equref{1111} into \equref{frame_energy}, all the cross-terms (where $k \neq m$) strictly vanish. The double summation effectively collapses into a single summation over the diagonal elements as $E_f = \sum_{k=0}^{MN-1} E_x g(0)$.
Since the equivalent pulse shaping filter is inherently normalized such that its energy is unity, i.e., $g(0)=1$, which corresponds to the diagonal elements of the FTN-induced Toeplitz correlation matrix $\mathbf{G}$, the total frame energy simplifies exactly to $E_f = MN E_x$. }
\par
\textcolor{red}{Given that the FTN frame duration is tightly compressed to $T_{\mathrm{frame}} \approx MN \alpha T_0$, the continuous-time average transmit power is consequently strictly constrained as
\begin{align} \label{P_avg}
	P_{\mathrm{avg}} = \frac{E_f}{T_{\mathrm{frame}}} = \frac{MN E_x}{MN \alpha T_0} = \frac{E_x}{\alpha T_0}.
\end{align}
This fundamental relationship confirms that to maintain a constant $P_{\mathrm{avg}}$ as the time compression factor $\alpha$ scales down, the discrete symbol energy $E_x$ must be proportionally reduced.  }

\vspace{-10pt}
	\section{}
	\textcolor{red}{
		Here, we provide a rigorous information-theoretic proof to justify the observation in Section V that the normalized SE decreases as the grid size $MN$ increases. 
		From \equref{spectral efficiency}, under the assumptions $\Delta f T_0 = 1$ and $E_f = MN E_x$, SE can be simplified as a normalized sum of subchannel capacities:
		\begin{equation}
			\eta = \frac{1}{\alpha} \left[ \frac{1}{MN} \sum_{n=0}^{MN-1} \log_2\left(1 + \frac{E_x \xi_n}{\sigma^2}\right) \right],
		\end{equation}
		where $\xi_n$ are the eigenvalues of the equivalent channel matrix, which are fundamentally governed by the FTN-induced colored noise covariance matrix $\mathbf{G}$. Let $f(x) = \log_2(1 + \frac{E_x}{\sigma^2}x)$ be the capacity function. Since $f''(x) = -\frac{(E_x/\sigma^2)^2}{\ln 2 (1 + E_x x/\sigma^2)^2} < 0$, $f(x)$ is a strictly concave function.  }
	\par
	\textcolor{red}{
		\textit{Lemma 1:} As the matrix dimension $MN$ increases, the dispersion (variance) of the eigenvalues $\{\xi_n\}$ of the Toeplitz matrix $\mathbf{G}$ strictly increases while their arithmetic mean remains constant.  }
	\par
	\textcolor{red}{
		\textit{Proof:} The arithmetic mean of the eigenvalues is given by the normalized trace:
		\begin{equation}
			\bar{\xi} = \frac{1}{MN} \sum_{n=0}^{MN-1} \xi_n = \frac{1}{MN} \text{Tr}(\mathbf{G}) = g_0,
		\end{equation}
		where $g_0$ is the self-correlation coefficient, as independent of $MN$. 
		The variance of the eigenvalues is:
		\begin{equation}
			Var(\xi) = \frac{\text{Tr}(\mathbf{G}_{MN}^2)}{MN} - g_0^2 = 2 \sum_{k=1}^{MN-1} \left( 1 - \frac{k}{MN} \right) |g_k|^2,
		\end{equation}
		where $g_k$ are the off-diagonal auto-correlation coefficients reflecting ISI. 
		For $k > 0$, $|g_k|^2 > 0$ due to FTN signaling. 
		It is evident that as $MN$ increases, the term $(1 - k/MN)$ increases for any fixed $k$, and the number of positive terms in the summation also grows. 
		Thus, $Var(\xi)$ is a strictly increasing function of $MN$, implying that the eigenvalues become more widely dispersed as the grid size expands.  }
\end{appendices}

\end{document}